\documentclass[10pt,titlepage]{article}

\usepackage{multirow}
\usepackage{amsfonts,amssymb,amsmath}

\usepackage{graphicx}

\usepackage[round,numbers,sort&compress]{natbib}
\newcommand{\invitro}{\textit{in vitro}}
\newcommand{\invivo}{\textit{in vivo}}
\newcommand{\ecut}{\ensuremath{E_{cut}}}
\newcommand{\waterfactor}{\ensuremath{w}}
\newcommand{\geometryfactor}{\ensuremath{\kappa}}

\newcommand{\flexindexaa}{\ensuremath{F_i}}
\newcommand{\etal}{et al.}

\begin{document}


\begin{titlepage}
\vspace*{-1.2in}


\vspace{3in}

\begin{center}
{\LARGE Stochastic kinetics of viral capsid assembly based on
detailed protein structures}

\vspace{.25in}

\Large{Martin Hemberg$^{\ast}$, Sophia N.\ Yaliraki$^{\dagger}$
and Mauricio Barahona$^{\ast}$}\footnote{\noindent Corresponding
author:  Dept. of Bioengineering, Imperial~College~London, South
Kensington Campus, London, SW7 2AZ (United Kingdom).
\mbox{  e-mail:  m.barahona@imperial.ac.uk} }

\vspace{.12in}

\large{Departments of $^{\ast}$Bioengineering and  $^{\dagger}$Chemistry, \\
  Imperial College London, United Kingdom}
\end{center}

\end{titlepage}

\abstract{We present a generic computational framework for the
simulation of viral capsid assembly which is quantitative and
specific. Starting from PDB files containing atomic coordinates,
the algorithm builds a coarse grained description of protein
oligomers based on graph rigidity. These reduced protein
descriptions are used in an extended Gillespie algorithm to
investigate the stochastic kinetics of the assembly process. The
association rates are obtained from a diffusive Smoluchowski
equation for rapid coagulation, modified to account for water
shielding and protein structure. The dissociation rates are
derived by interpreting the splitting of oligomers as a process of
graph partitioning akin to the escape from a multidimensional
well. This modular framework is quantitative yet computationally
tractable, with a small number of physically motivated parameters.
The methodology is illustrated using two different viruses which
are shown to follow quantitatively different assembly pathways. We
also show how in this model the quasi-stationary kinetics of
assembly can be described as a Markovian cascading process in
which only a few intermediates and a small proportion of pathways
are present. The observed pathways and intermediates can be
related \emph{a posteriori} to structural and energetic properties
of the capsid oligomers.}


\clearpage

\section*{Introduction}

Viruses are the cause of some of the deadliest diseases today.  In
fact, the lethality of viruses emanates from their simplicity; as the
ultimate non-autonomous parasites, viruses cannot replicate without a
host cell and are therefore immune to standard anti-bacterial drugs.
Basically, a virus consists of two components: genetic material (DNA
or RNA) and a protective protein shell, the \textit{capsid}.  In a
self-referencing loop, the viral nucleic acids encode the proteins
that form the viral capsid. Once the virus penetrates a host cell, it
hijacks the cellular machinery of the host and uses it to replicate
the viral genome and to express the viral protein(s), which then
assemble into capsids. As a result, the infected cell acts as a
replicator of new viruses instead of performing its normal
tasks~\citep{Alberts:2002}.

\begin{figure}
  \begin{center}
 \includegraphics[width = 3.25in]{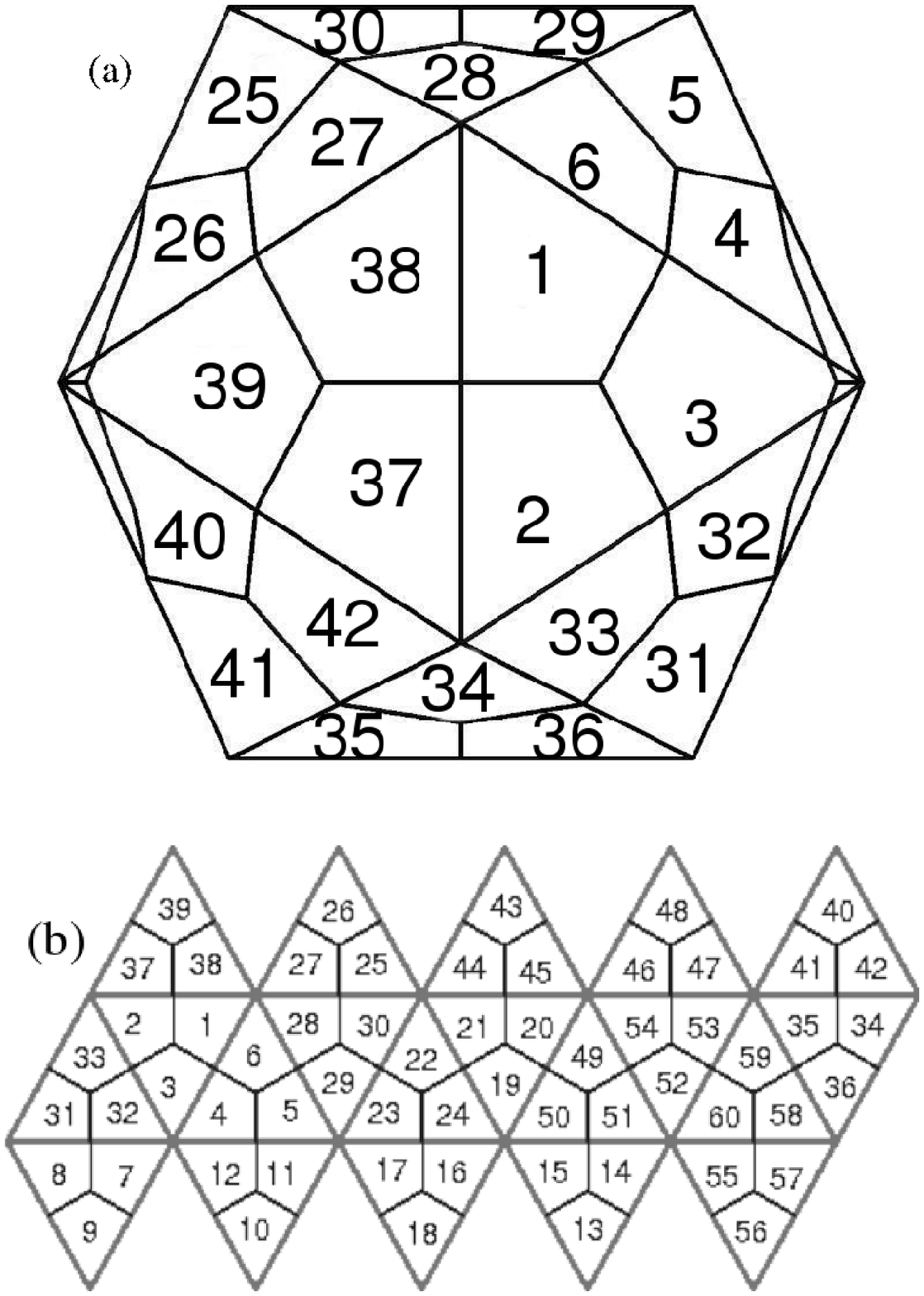}
    \caption{}
    \label{fig:1stmInterfaces}
  \end{center}
\end{figure}

Another remarkable feature of viruses is that capsids are commonly
quasi-spherical with icosahedral
symmetry~\citep{Alberts:2002,Caspar:1962}. Although other viral
structures, such as cigar shaped and partial sheets, are possible,
we restrict our investigation to icosahedral capsids. Because
encoding a large protein to envelop the whole genome is not
physically realizable, identical copies of the same protein are
used in a symmetric arrangement. Therefore, symmetry is used to
economize the number of distinct proteins encoded in the viral
genomes. This was formalized beautifully in the classic theory of
\emph{quasi-equivalence}~\citep{Caspar:1962,Zlotnick:2004}, which
broadly predicts the manner in which identical asymmetric protein
units can be used to form a symmetric capsid. Quasi-equivalent
viruses are characterized by their $T$--number, the number of
proteins in each asymmetric unit~(Fig.~\ref{fig:1stmInterfaces}).
This leads to icosahedral capsids with $60 \, T$ proteins, where
geometrical constraints dictate that $T = h^2 + hk + k^2$, with
$h$ and $k$ non-negative integers. Clearly, viral capsids with
larger $T$--values enclose a larger volume while maintaining
icosahedral symmetry.

The assembly of the capsid, a crucial step in the virus life
cycle, could provide an opportunity to interfere with the process
of virus replication~\citep{Zlotnick:2003}. However, although
there is a wealth of structural capsid data from X-ray
crystallography and cryo-electron microscopy, the assembly
pathways remain largely uncharted. It is known that inside the
cell the capsid is assembled around the virus genome (DNA or RNA)
with only limited or no assistance from other
bio-molecules~\citep{Alberts:2002}. Even more remarkable, for some
viruses self-assembly can take place \invitro\ , in the absence of
the genome and outside the cellular environment, and still lead to
stable capsids that are indistinguishable from those created
\invivo. The role of the genome in the assembly process is not
fully clarified and it may well be that \invivo\ and \invitro\
assemblies follow different routes~\citep{McPherson:2005}.

Because detailed experimental data on assembly routes is at
present difficult to
obtain~\citep{Schwartz:1998,Larson:2001,Fox:1998,Baker:1999,
McPherson:2005}, modelling and simulation approaches have come to
play an important role in the understanding of this process. In
particular, one would like to identify the pathways by which the
oligomers combine to form the final capsid and the factors that
can influence the process. Previous theoretical work has
approached different aspects of the assembly process using a
variety of techniques: from dynamic to static models, both
microscopic and macroscopic~\citep{Schwartz:1998, Berger:1994,
Rapaport:2004, Horton:1992, Reddy:1998, Sitharam:2004,
Hespenheide:2004, Endres:2002, Zlotnick:1999, Zlotnick:2003,
Ceres:2002, Kegel:2004}.

Ideally, a fully dynamic view of the assembly process could be
achieved by performing molecular dynamics (MD) simulations with a
full atomic description of the proteins in aqueous solution.
However, the computational cost of full MD restricts its
applicability to simplistic models of proteins---essentially,
balls with sticky pods under Brownian motion. Schwartz and
Berger~\citep{Berger:1994, Schwartz:1998} performed such a
dynamical simulation of capsid formation, where they showed that
the assembly can be completed using only \emph{local} information
in the incomplete capsid. Recently, \citet{Rapaport:2004} has
presented more realistic MD results that capture some of the
salient features of a generic virus self-assembly process but
still lacking the necessary detail to investigate specific
viruses.

Quantitative results for specific viruses can only be obtained
through the use of a more detailed protein model. However, it is
currently infeasible to simulate explicit dynamics of such a large
ensemble of hydrated proteins due to the size and complexity of
the units. This has led to microscopic approaches in which the
partially completed capsid is investigated as it is assembled
\emph{quasi-statically}. The assumption here is that the relative
positions (and thus, the interactions and energies) in the
incomplete capsid are identical to those found in the complete
capsid. However, this is itself a computationally hard problem due
to the combinatorial number of assembly pathways.
\citet{Horton:1992} were the first to use combinatorial
optimization to find substructures with the most favorable
association energies. This scheme was further extended
by~\citet{Reddy:1998} with a more refined method for calculating
the energies. Beyond purely energetic considerations, structural
concepts have been used to characterize protein assemblies:
\citet{Sitharam:2004} have used combinatorial and computational
algebra to create models based on static geometric and tensegrity
constraints, while \citet{Hespenheide:2004} have investigated
rigid protein assemblies as likely candidates to be long-lived.

Alternatively, other theoretical studies have concentrated on more
macroscopic approaches. Some studies have focused on the static
mechanical structure of the full capsid rather than the dynamics of
the assembly~\citep{Lidmar:2003, Marzec:1993, Tarnai:1995}.  Recent
work of Bruinsma \etal ~\citep{Bruinsma:2003, Zandi:2004} (see also
~\citep{McPherson:2005} for more qualitative ideas) is based on
statistical mechanics calculations of free energies that take into
account the curvature of the capsid. Finally, the macroscopic kinetic
approach pursued by the group led by Zlotnick~\citep{Endres:2002,
Zlotnick:1999, Zlotnick:2003, Ceres:2002} (see
also~\citep{Kegel:2004}) describes capsid assembly through empirical,
law of mass action differential equations for the concentration of the
different oligomers.  However, although the results can be related to
bulk concentration measurements, this kinetic approach is still unable
to provide information about microscopic pathways. In recent work,
\citet{Endres:2005} have concluded that only a few out of the
combinatorially many intermediates play any role and that these cannot
be predicted by considering minimal energy configurations alone.

In this paper, we develop a modelling framework that incorporates
atomic detail of proteins into an explicit implementation of the
kinetics of capsid assembly as a stochastic process. Our model
starts from atomic descriptions of the protein oligomers,
available from databases such as VIPER~\citep{Reddy:2001}, and
simplifies the representation through a reduction of the degrees
of freedom based on graph rigidity measures with the aid of the
software FIRST~\citep{Jacobs:2001}.  These reduced oligomer
descriptions are used to simulate stochastically the process of
capsid formation, without allowing for malformed structures,
through an extended Gillespie algorithm~\citep{Gillespie:1977}.
Our scheme includes both diffusive association and dissociation
reactions whose reaction rates are derived using the reduced graph
representations. Although our algorithm does not implement
dynamics explicitly, it provides the stochastic time evolution of
the system and the quasi-steady oligomer distribution. This
information can be analyzed to infer which pathways are important
in the assembly of specific viruses and the role that protein
structure and chemical environment play in the assembly process.

\section*{Reduced protein descriptions from full atomic models} \label{sec:proteinModel}

In order to incorporate sufficient molecular detail, our
computational framework starts from the detailed atomic structure
of proteins as determined by crystallographic experiments. An
invaluable resource is the database VIPER~\citep{Reddy:2001},
which provides protein structures, transformation matrices, maps
for adjacent proteins and binding energies for a large number of
viruses. This full atom description of the protein oligomers needs
to be simplified to make it tractable for computational purposes.
The basic physical idea underlying our simplified protein model is
the assumption that rigid substructures will effectively move as a
block. This implies a reduction in the number of degrees of
freedom and, consequently, in the effective size of the problem.

The initial step is the addition of hydrogen atoms to the PDB
structure using the software WHAT IF~\citep{Vriend:1990}. We then
characterize the full atom structure of each oligomer with FIRST, a
computational tool for the analysis of proteins developed
by~\citet{Jacobs:2001}. FIRST uses standard potentials to identify
covalent and hydrogen bonds, salt bridges and hydrophobic tethers in
the structure, and represents the protein as a bond bending
network. This graph representation, where nodes are atoms and edges
indicate constraints introduced by bonds, is then analyzed with a
computationally efficient algorithm (the pebble game) to identify
flexible(underconstrained) and rigid (overconstrained)
regions~\citep{Jacobs:1998}. FIRST also calculates the energies for
all the bonds in the protein network.

\begin{figure}
  \begin{center}
  \centerline{\includegraphics[width = 6.5in]{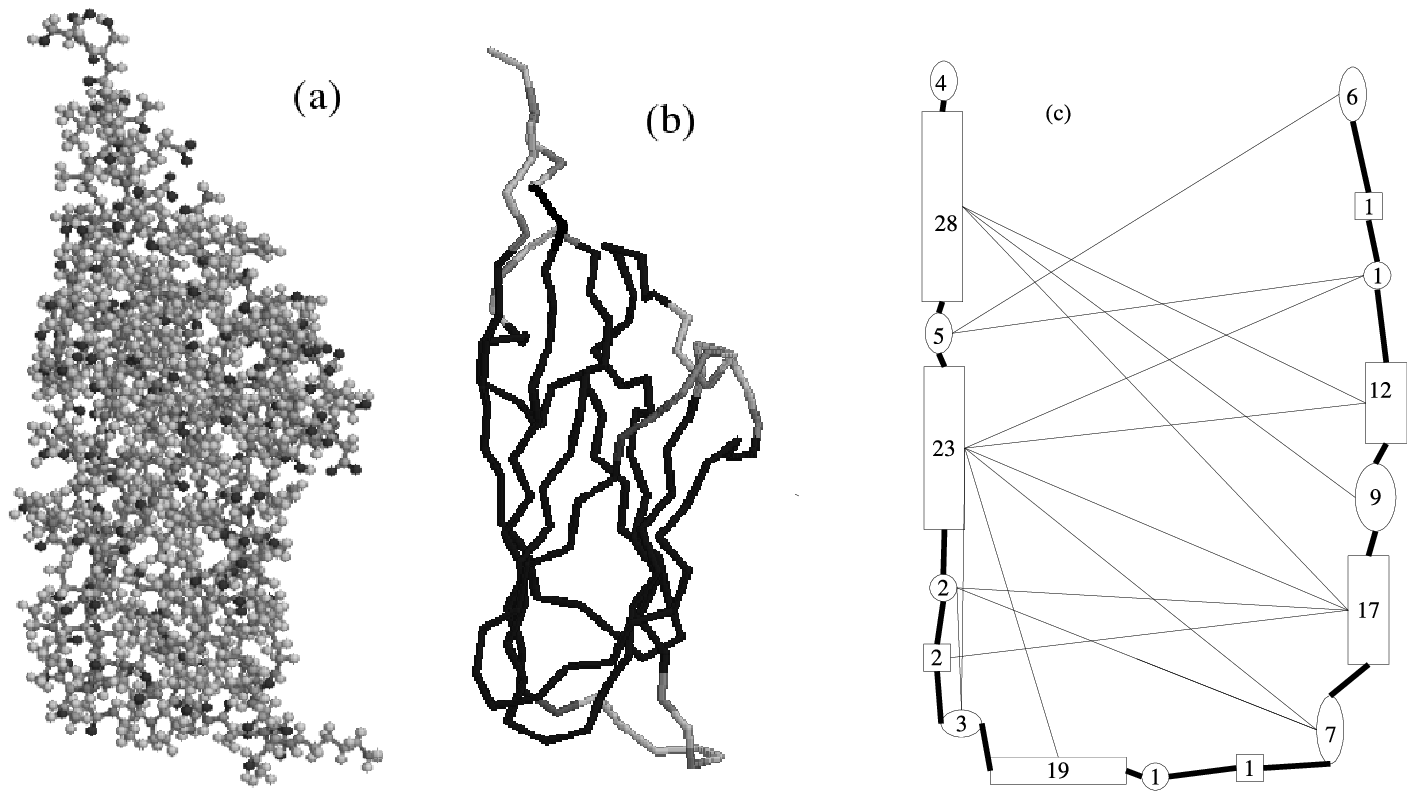}}
    \vspace{-3in}
    \caption{}
    \label{fig:1stmFlexibility}
  \end{center}
\end{figure}

\begin{figure}
\vspace{-2in}
  \begin{center}
  \centerline{\includegraphics[width = 8in]{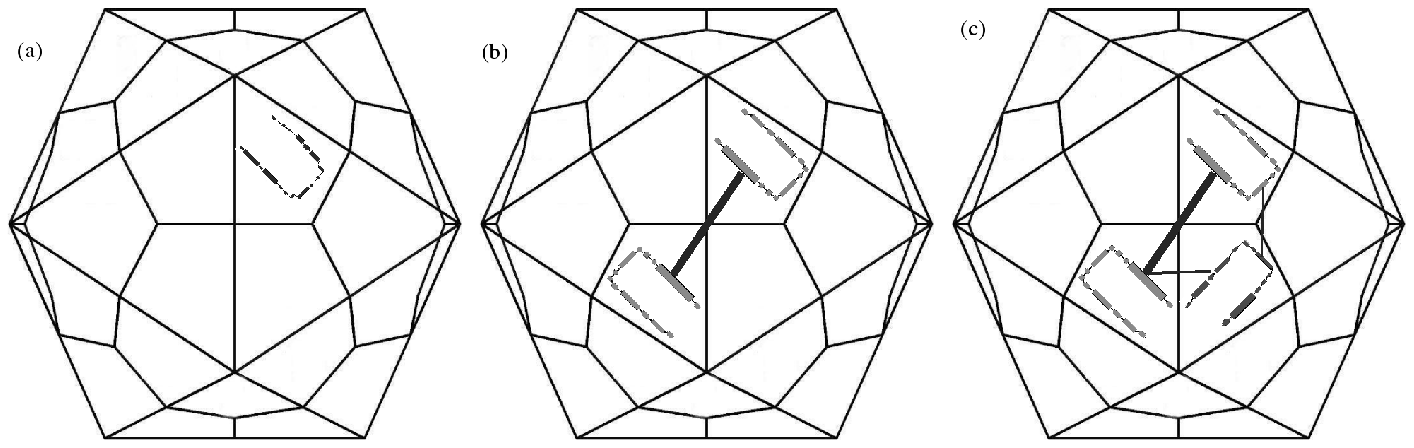}}
  \vspace{-4in}
  \caption{}
  \label{fig:1stmRigidityChange}
  \end{center}
\end{figure}

The output from FIRST can be used to produce a flexibility index
\flexindexaa\ for each amino acid~\citep{Jacobs:2001}. When
\flexindexaa\ $\leq 0$, the amino acid is overconstrained, and
therefore rigid; when $\flexindexaa >0$ the amino acid is floppy
(underconstrained). We then group adjacent residues with the same
binary rigidity into rigid and floppy \emph{domains}. As shown in
Fig.~\ref{fig:1stmFlexibility}, a protein typically consists of long,
rigid domains separated by short, floppy hinge segments. It is
important to point out that because graph rigidity is a nonlinear
property, the rigidity of a protein may change as the aggregation
proceeds, even though none of the atoms have moved relative to its
neighbors. When two proteins bind, new constraints are added to the
graph, usually leading to a more rigid network (see
Fig.~\ref{fig:1stmRigidityChange}). The procedure outlined in this
section amounts to a significant coarse-graining of the model: it
starts from a full description (PDB file) with several thousand atomic
coordinates for each protein and it outputs a representation
consisting of a few rigid blocks (on the order of a few tens per
monomer). It is this reduced representation (illustrated in
Fig.~\ref{fig:1stmFlexibility}) that we use to implement the
stochastic kinetics of self-assembly.

\section*{Stochastic kinetics of capsid assembly} \label{sec:StochasticModel}

Studying the time evolution of the assembly process by integrating the
equations of motion is computationally infeasible even for reduced
representations like those described above. There are two main
obstacles for the implementation of a fully dynamical approach: first,
the combinatorial explosion of the number of intermediates for large
aggregates of proteins---a problem that cannot be overcome by sheer
computational power and that must be addressed at the modelling stage;
second, the lack of tested and rigorous coarse-grained potentials for
explicit dynamics of reduced protein models, especially when diffusion
plays a significant role. To circumvent these problems, we consider
instead the stochastic kinetics of the assembly process through an
extended version of Gillespie's stochastic algorithm in which we
consider dissociation and association events modulated by diffusion.

\begin{figure}
  \vspace{-2in}
  \begin{center}
  \centerline{\includegraphics[width = 6.5in]{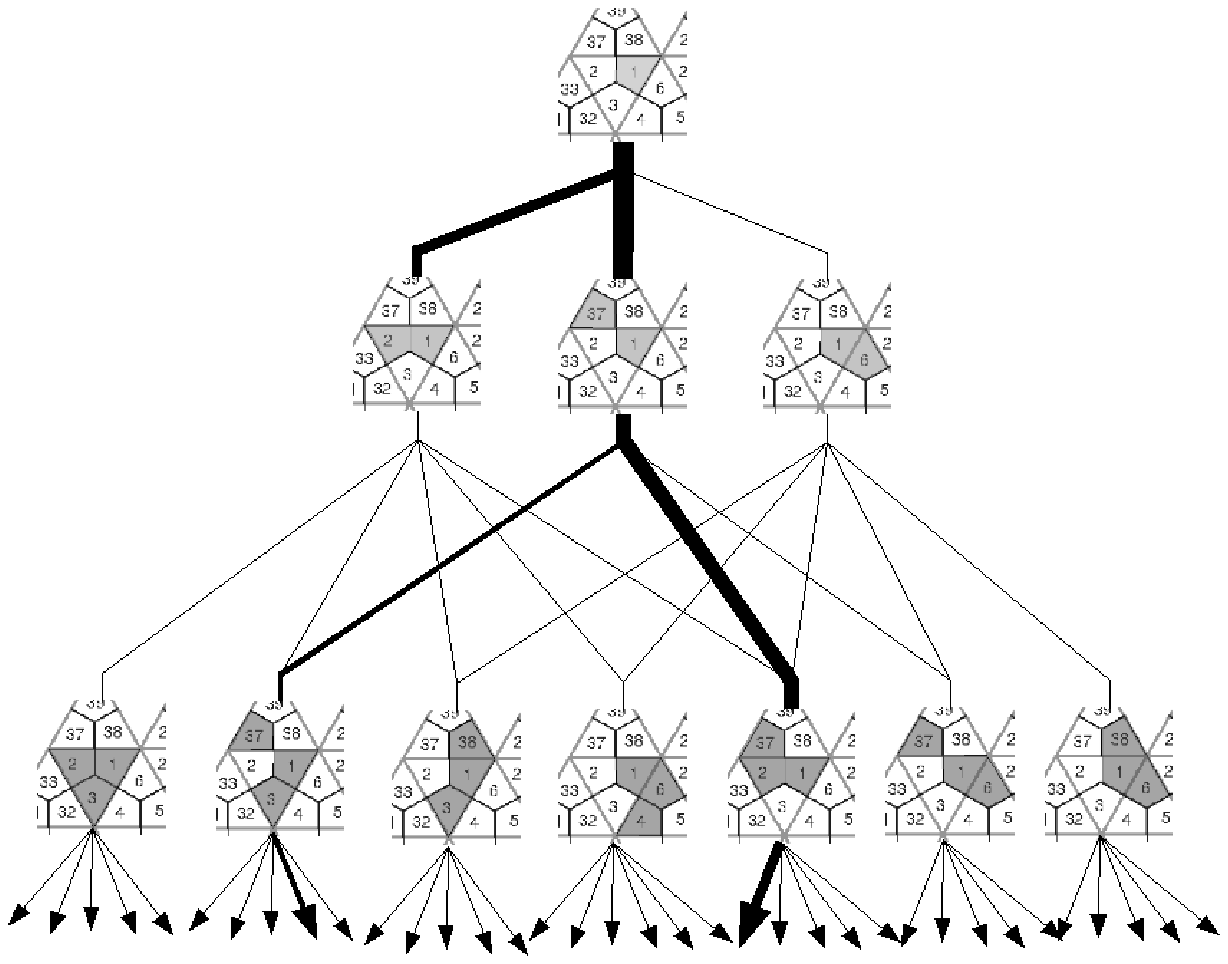}}
        \vspace{-2in}
   \caption{}
    \label{fig:1stmAssemblyTree}
  \end{center}
\end{figure}

Gillespie's classic algorithm~\citep{Gillespie:1976, Gillespie:1977}
was introduced in 1976 as a computational tool for the stochastic
simulation of chemical reactions. Recently, Gillespie's algorithm has
had a vigorous revival due to its relevance to many biological
systems, where only small numbers of molecules are present. The
theoretical basis for a stochastic formulation of chemical reactions
is the chemical master equation which describes the probability that a
given event (or no event) takes place over an infinitesimal time
interval~\citep{vanKampen:1992}. Unfortunately, the master equation is
not solvable explicitly for systems involving more than a few
different molecules and reactions. Gillespie's algorithm addresses
this numerically and provides an exact procedure for a Monte Carlo
simulation of a system of reacting molecules. As is obvious in
Fig.~\ref{fig:1stmAssemblyTree}, the complexity of the pathways
increases combinatorially with the size of the oligomers. The
propensity of each reaction is a product of a combinatorial factor,
dependent on the number of reactant molecules available for the
reaction, and a rate constant, dependent on properties (such as size,
velocity and mass) of the molecules involved in the
reaction~\citep{Turner:2004}.

\subsection*{Association events}

During capsid assembly, there are association and dissociation
events. The association events are elementary (bi-molecular)
`reactions' in which two oligomers collide to form a new complex.
The association process of structured molecules in solution can be
modelled as a succession of two independent processes: first, two
oligomers must meet through a diffusive process; next, they must
overcome a barrier in order to aggregate and reach the final bound
state~\citep{Selzer:2001, Janin:1997}. In its standard form, the
Gillespie algorithm assumes that the reactants are dilute,
perfectly-mixed, structure-less molecules in vacuum. This is
obviously not a good approximation in our case, and we have
extended the algorithm to take into account diffusion, the
influence of water, and geometric and entropic factors. Our
approach is simpler than the explicit stochastic simulation of the
spatio-temporal reaction-diffusion process using computationally
intensive voxel models~\citep{Stundzia:1996, Bernstein:2005} yet
it captures the relevant physical features.

To account for the diffusive rate, we use concepts from
Smoluchowski's theory of rapid coagulation~\citep{Drake:1972}. In
its simplest form, this theory was developed for spherical
colloidal particles and hence needs to be corrected when applied
to protein aggregates with specific geometry and binding
sites~\citep{Camacho:1999,Selzer:2001}. It can be shown that the
modified Smoluchowski rate is:
\begin{equation}
  \label{eq:smoluchowski}
  k_{ij}^{\rm assoc} =  k_{ij}^{\rm hs} \geometryfactor =  \left (4\pi
  D_{ij}R_{ij}n_in_j \right )
   \, \geometryfactor,
\end{equation}
where $k_{ij}^{\rm hs}$ is the Smoluchowski diffusive rate for
hard spheres. Here, $n_i$ and $n_j$ are the unit concentrations of
particle types $i$ and $j$, and the diffusivity $D_{ij}R_{ij} =
D_1r_1(r_i^{-1} + r_j^{-1})(r_i + r_j)$ is related to $D_1$ and
$r_1$, the diffusion coefficient and radius of the monomer, and to
$r_i$ and $r_j$, the radii of particle types $i$ and $j$. Based on
a simple scaling geometric argument valid for disc-shaped
oligomers, it can be assumed that the radius increases as the
square root of the number of monomers. The dimensionless parameter
\geometryfactor \ is a form factor, which reflects the probability
that a collision between two oligomers will result in the
formation of a complex.  It accounts for the fact that the
proteins will attach at a lower rate than homogenously `sticky'
particles due to their geometry and specific binding sites. It can
also be interpreted as a generic entropic barrier that needs to be
surmounted for association~\citep{Janin:1997}.

The aggregation of oligomers can occur in a number of different
ways with different association energies $E^{(a)}_{ij}$ for the
specific pairings (see Table~\ref{tab:interfaceEnergies}). When
forming a new oligomer we assume the proteins to be at the
positions that they attain in the complete capsid. This means that
our model does not account for malformed capsids. Neither does it
include the maturation or conformational changes that are known to
occur in some viruses. To model the fact that oligomers with large
negative association energies are more likely to be formed, we
multiply the rate in Eq.~\ref{eq:smoluchowski} by a Boltzmann
factor $\exp (\waterfactor E^{(a)}_{ij}/k_B T )$, where $k_B$ is
the Boltzmann constant, $T$ is the temperature and $E^{(a)}_{ij}$
is the association energy. The association energy is modified to
include the effect of water shielding. Because the protein is
surrounded by water, the effective energy of inter-protein
hydrogen bonds is reduced, since there exists the alternative of
forming bonds with water molecules instead. It is important to
note that both the form factor \geometryfactor\ and the water
shielding factor \waterfactor\ can be estimated from experiments
or molecular dynamics simulations~\citep{Selzer:2001,
Northrup:1992, Sheu:2003}.

The energies can be obtained from different sources but, for
simplicity, we have used throughout this paper the energies as
calculated by FIRST. We remark, however, that our algorithm is
modular and more sophisticated energy calculations could be easily
incorporated into our computational framework, e.g., CHARMM
energies from VIPER~\citep{Reddy:2001, Brooks:1983}. For
completeness, we have carried out a comparison between the
association energies from FIRST and VIPER. We have checked that
although the energies can differ significantly in absolute numbers
(as shown in Table \ref{tab:interfaceEnergies}), both the ordering
of the bond strengths and the localization of the bonds are
broadly consistent between FIRST and VIPER.

\subsection*{Dissociation events} \label{sec:dissociation}

In addition to aggregating, oligomers can also break up into
smaller units with a dissociation rate which is an indication of
the longevity of an oligomer. The propensity of a dissociation
event depends on the energy required to break the bonds that hold
the oligomer together, but is also related to the redistribution
of energy into the internal modes of the oligomer. It is in this
context that our reduced description of protein oligomers becomes
most helpful.

We base our modelling of the dissociation process on
transition-state theory as applied to the escape from a
multi-dimensional well. In this framework, the escape rate from a
well with $N$ vibrational degrees of freedom is given
by~\citep{Hanggi:1990}
\begin{equation}
  \label{eq:multidimensionEscapeRate}
  k^{\rm dissoc} = \frac{1}{2 \pi} \frac{\prod_{i=0}^N \lambda_i^{(0)}}
  {\prod_{i=1}^N \lambda_i^{(b)}}\exp \left (-E^{(b)}/k_B T \right ),
\end{equation}
where $\lambda_i^{(0)}$ are the eigenfrequencies at the bottom of
the well and $\lambda_i^{(b)}$ are the eigenfrequencies when the
particle is at the point of escape (i.e., at the top of the
barrier of height $E^{(b)}$). The generic
Eq.~\ref{eq:multidimensionEscapeRate} can be related to the
reduced protein model introduced in the previous section. If we
view the oligomer as a harmonic network, where each domain is
treated as a point mass and the bonds connecting domains as linear
springs, then the original oligomer represents a local minimum in
the energy landscape and escape from this well represents the
physical process of splitting the oligomer.

The eigenfrequencies of the system at equilibrium,
$\lambda_i^{(0)}$, are obtained by diagonalizing the system
$M\ddot{x} + K x = 0$, where $M$ is the diagonal matrix of domain
masses and $K$ is the weighted Laplacian matrix of the graph. Each
weight $K_{ij}$ is the stiffness of the bond connecting domain $i$
and $j$ obtained from Hooke's law, $K_{ij} = 2E_{ij}/x_{ij}^2$,
with $E_{ij}$ being the energy of the bond and $x_{ij}$ the
equilibrium distance of the bond. The diagonal elements of the
stiffness matrix, $K_{ii}$, are given by the condition that the
sum of the elements in each row is zero. In our reduced network,
there are two types of bonds to be included in the analysis:
hydrogen bonds and covalent bonds.  The energies are provided by
FIRST: hydrogen bonds are of the order of $-5$ kcal/mol once they
have been multiplied by the water shielding pre-factor
\waterfactor, and we assume the covalent bonds to be $-74$
kcal/mol, a value close to the typical energies of C--N and C--C
bonds. (Note that this energy is not multiplied by \waterfactor\
since there is no option for the covalent bonds to form bonds with
the surrounding water molecules.) If two domains are linked by
both hydrogen bonds and covalent bonds, only covalent bonds are
considered since they are an order of magnitude stronger.

To obtain the dissociation rate in
Eq.~\ref{eq:multidimensionEscapeRate} for a given split, we first
calculate the eigenfrequencies of the original system
$\lambda_i^{(0)}$ via the generalized eigenvalue problem of the
unmodified network. The eigenvalues of the system at the barrier,
$\lambda_i^{(b)}$, and the barrier height, $E^{(b)}$, are obtained by
examining the possible partitions of the graph. A given partition is
characterized by the minimal set of edges that is required to split it
into two subgraphs. The total energy of the removed edges is equal to
$E^{(b)}$ and the $\lambda_i^{(b)}$ are obtained as the generalized
eigenvalues of the partitioned graph.  Indeed, when the graph is
partitioned one eigenvalue becomes zero, which explains why the
numerator and denominator do not run over the same indices. For most
oligomers, the most favorable splits have an eigenmode ratio of about
$10^{-2}$, although this ratio can be up to five orders of magnitude
larger in some cases. Similarly to the association events, each split
is then considered within our Gillespie algorithm as an independent
event with its own characteristic propensity with rate $k^{\rm
dissoc}$ given by Eq.~\ref{eq:multidimensionEscapeRate}.

Clearly, there are many ways of splitting an oligomer. For example,
the trimer in Fig.~\ref{fig:1stmRigidityChange}\textit{c} can split in
three different ways with different propensities. Since the total
number of possible splits grows combinatorially with the number of
domains in the oligomer, we have reduced the complexity by imposing
two constraints on the partitions: only hydrogen bonds are allowed to
break, and only bi-partitions,(i.e., splits into two fragments) are
considered. The latter is not an extreme assumption as splits
resulting in more fragments can be composed of a number of subsequent
bi-partitions. Under these restrictions, and due to the sparsity of
the inter-monomer connections in the icosahedral lattice, the number
of partitions grows sub-exponentially, the exact rate depending on the
topology of the oligomer. The eigenvalue calculation for the
dissociation events is the most time consuming step in our
simulations. To speed up the calculations, we have devised a data
structure that stores the results of known events for use in
subsequent runs.

\section*{Results of the simulations} \label{sec:results}

Our modelling framework is generic and can be applied to any
icosahedral virus. In this section, we illustrate the output of
the current version of the program with two small plant viruses:
the $T=1$ virus Satellite Panicum Mosaic Virus (SPMV, PDB code:
1stm) and the $T=3$ virus Southern Bean Mosaic Virus (SBMV, PDB
code: 4sbv). Interestingly, SBMV is known to be capable of
self-assembly \invitro\ \citep{Savithri:1983} whereas SPMV is not.
Before we present some numerics, we make two technical points
regarding the simulations.

One advantage of our computational model is that it has relatively
few, physically meaningful parameters.
Table~\ref{tab:parameterValues} presents a summary of the
parameters: (i) the temperature and concentration; (ii) the
average radius and diffusion coefficient of a monomer in order to
calculate the diffusion rate in Eq.~\ref{eq:smoluchowski}; and
(iii) the bond constants used to derive the eigenfrequencies for
the dissociation rate in Eq.~\ref{eq:multidimensionEscapeRate}.
All these quantities are directly related to physical variables.
There are three additional parameters that, although physically
motivated, are of a more conceptual nature. First, \ecut\ is an
input parameter for the software FIRST that specifies the cut-off
energy for a hydrogen bond to exist. This can be loosely related
to temperature and under standard conditions it is approximately
$-0.7$ kcal/mol~\citep{Jacobs:2001}. Second, the fact that
proteins are surrounded by water means that the effective strength
of the hydrogen bonds is reduced by a factor \waterfactor\ which
has been estimated to be 10--25\% using detailed MD
calculations~\citep{Sheu:2003}. This parameter is related to pH
and to the ionic strength of the solution. Third, the form factor
\geometryfactor\ used in the modified Smoluchowski equation
(Eq.~\ref{eq:smoluchowski}) has been estimated to be in the range
$10^{-3}-10^{-5}$ through computer simulations of diffusing
proteins~\citep{Selzer:2001, Northrup:1992}. This parameter is
related to the specific geometry and docking of the
oligomers.

The second technical point refers to size limitations in the software
used. Our current implementation uses version 3.1 of the software
FIRST, which is limited to analyzing protein structures with a maximum
of $75,000$ atoms~\citep{Jacobs:2001}. This limitation is not
intrinsic to the method (only to version 3.1) and future releases will
extend its capabilities. This effectively means that at present we do
not investigate \emph{dissociation} paths for oligomers larger than 20
proteins even if the computations are fast. Therefore, our full
simulations (including both association and dissociation propensities)
are run up to the formation of oligomers of size 20. However, we will
also present simulations of the completion of the full capsid obtained
from runs with association paths alone, which do not rely on the use
of FIRST.

The starting point for the simulations is a state where all
proteins are present as monomers. The system then evolves towards
aggregation into larger units. There is an initial transient
during which a large amount of reactions involving monomers take
place. Relatively quickly, the concentration of a few key
oligomers builds up and the system then settles into a
quasi-steady state in which the concentration of the different
oligomers remains relatively constant---except, of course, for
monomers and `completed' capsids (size larger than 20).
Effectively, monomers are transformed into capsids via restricted
pathways that do not alter significantly the average
concentrations of the intermediates. We explore and characterize
this cascading process in what follows.

\subsection*{The quasi-steady solution}

\begin{figure}
  \begin{center}
    \includegraphics[width=3.25in]{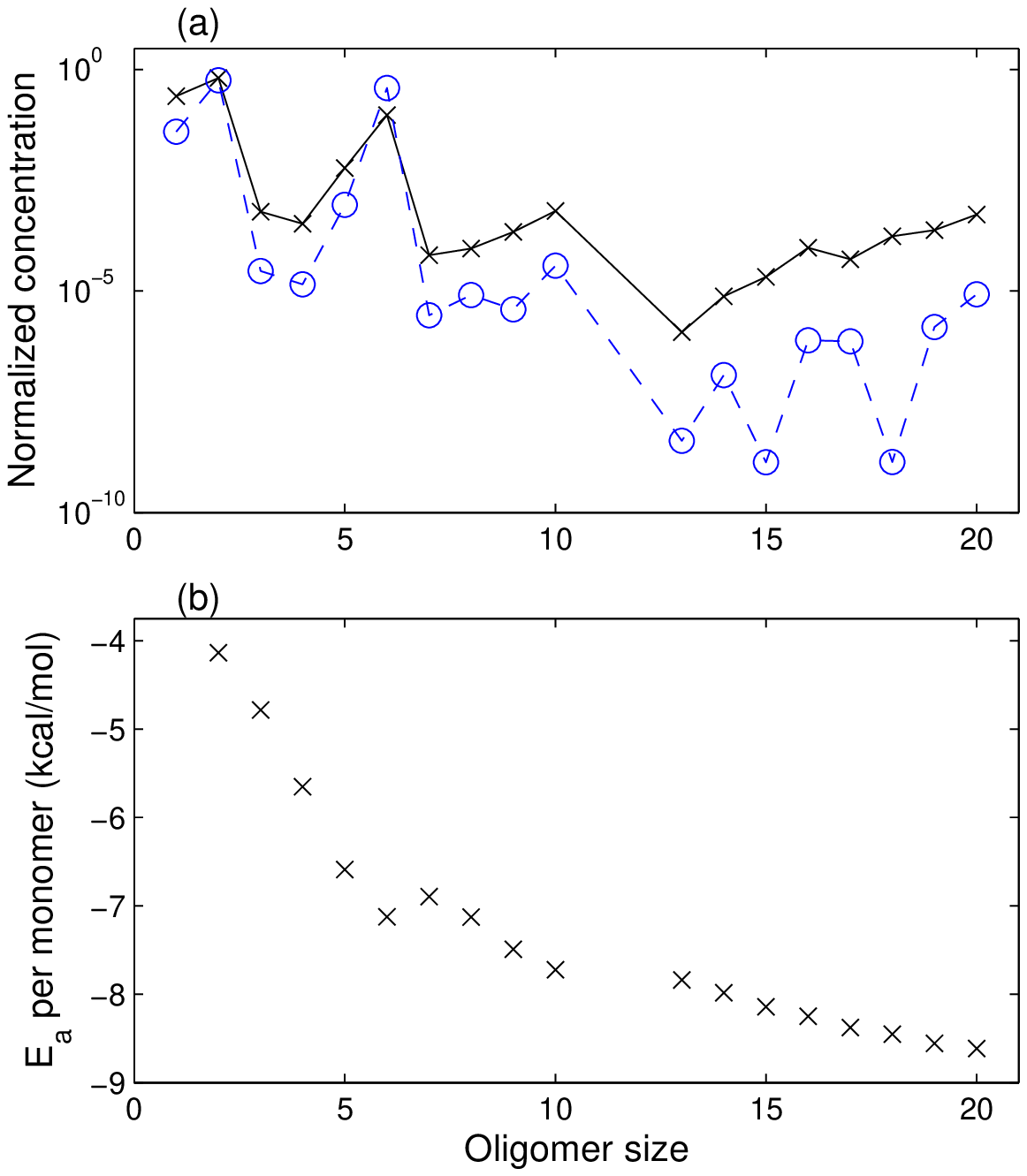}
    \caption{}
    \label{fig:1stmConcEA}
  \end{center}
\end{figure}

We first illustrate some of the results through the analysis of the
quasi-steady state in the assembly of the $T=1$ SPMV virus
(1stm). Each simulation starts with $1,000$ monomers. To eliminate the
transient, we do not collect statistics until the first oligomer of
size larger than 20 is formed. At this point, we consider the system
past the transient state, we remove the large oligomer and we record
the time-weighted concentration average of all oligomers until the
next $> 20$-mer is formed. We repeat this procedure $1,000$ times and
average the results, which are presented in
Fig.~\ref{fig:1stmConcEA}. It is important to note that we have also
run simulations where starting from an empty system, we add monomers
at a constant rate and remove oligomers larger than size 20. Once this
open system reaches a quasi-steady state, we have checked that it
behaves in the same way on average as the one starting from a fixed
number of monomers.

Inspecting the simulations, we find that there are very few distinct
oligomers with a significant presence throughout the process
(Fig.~\ref{fig:1stmConcEA}). Only monomers, dimers and hexamers are
present in any significant concentration during the formation of the
capsid. The concentrations of all other oligomers are negligible. A
similar conclusion was also reached
by~\citet{Endres:2005}. Interestingly, this is not just the result of
the difference in the association energies; Fig.~\ref{fig:1stmConcEA}
shows that all oligomers have similar association energies (per
monomer).

The simulation data also yield information about the processes
governing the kinetics of the system. Oligomers larger than hexamers
are quite rare and as soon as one is created it tends to participate
in a series of rapid reactions leading to a $>20$--oligomer. This
cascading behavior emerges because large intermediates tend to follow
from favorable association energies and also tend to be stable with
respect to dissociation. This view of the assembly as a cascading
process is in good agreement with other dynamical
simulations~\citep{Endres:2002,Rapaport:2004}.

Oligomers with a significant concentration (monomers, dimers and
hexamers) tend to fluctuate around a mean value. On the other hand,
oligomers with negligible concentrations are not present most of the
time and they react and disappear quickly when present. These two
types of behavior are illustrated in
Fig.~\ref{fig:1stmPoissonGaussian} where we plot the average
probability distributions of the concentration of dimers, tetramers
and hexamers at quasi-steady state. The distribution of dimers is
`Gaussian-like' around a high concentration while tetramers show the
characteristics of a `Poisson-like' distribution. Hexamers display
less clear features.  Indeed, although all the underlying elementary
stochastic processes of aggregation and dissociation are Markovian,
the structure of the kinetic network leads to a variety of
quasi-steady distributions for the different intermediates. In the
Discussion, we provide a simple theoretical argument of how these
distinct behaviors emerge.

\begin{figure}
  \begin{center}
   \centerline{\includegraphics[width = 5in]{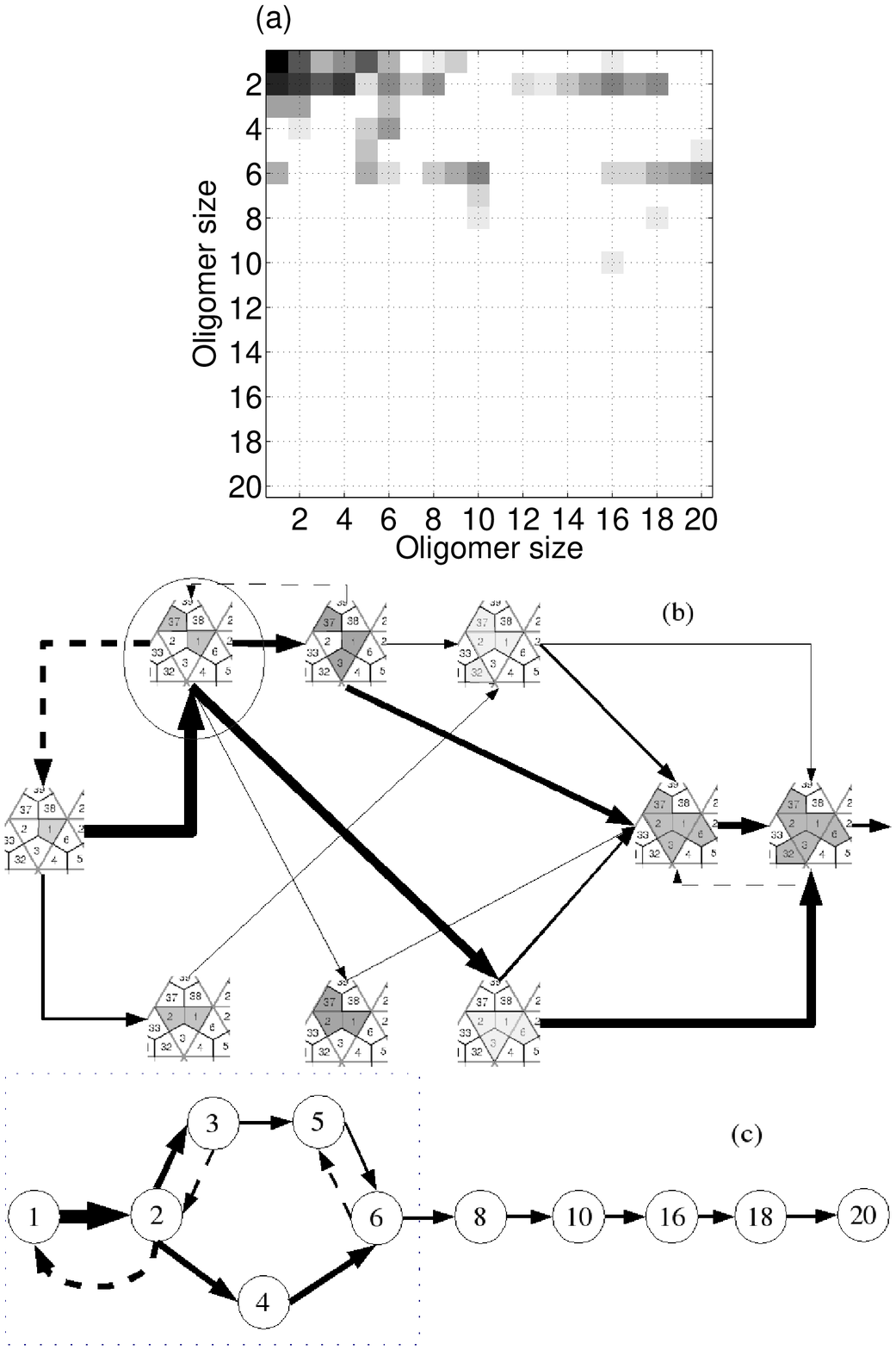}}
    \caption{}
    \label{fig:1stmReactionCartoon}
  \end{center}
\end{figure}

The simulations of the assembly of 1stm can be used to extract
further details about the pathways in use in the network of
reactions. To make this more explicit, we calculate the average
transition probability of association and dissociation events as
derived from the numerics. These probabilities form a transition
matrix, which we present in
Fig.~\ref{fig:1stmReactionCartoon}\textit{a} as a heat map. The
upper triangular section of the matrix corresponds to association
processes while the lower triangular section corresponds to
dissociation reactions. Note that virtually all the dissociation
events are confined to the small oligomers. One of the reasons
that small oligomers are easier to split is that they have fewer
inter-monomer bonds per protein. This can be understood from
Eq.~\ref{eq:multidimensionEscapeRate} where the Boltzmann factor
has a large impact on the dissociation rate.

Using these data, we show in
Fig.~\ref{fig:1stmReactionCartoon}\textit{b,c} that the assembly
proceeds via a few pathways which thread through the combinatorially
complex association/dissociation tree shown in
Fig.~\ref{fig:1stmAssemblyTree}. These reactions lead to significant
quasi-steady concentrations only for monomers, dimers, hexamers,
10--mers, 16--mers and 20--mers. A mere inspection of the binding
energies prior to the simulations, would not lead to this outcome
although it can be understood \textit{a posteriori} in terms of the
properties of the oligomers. For instance, almost all the 1stm dimers
formed correspond to the dimer circled in
Fig.~\ref{fig:1stmReactionCartoon}\emph{b}, which has a bond with
2--fold icosahedral symmetry in the full capsid (see
Fig.~\ref{fig:1stmInterfaces} and
Table~\ref{tab:interfaceEnergies}). Since the other bonds involved in
dimers are significantly weaker there will hardly be any other dimers
present. The predominance of this dimer has consequences: the dominant
hexamer can in turn be viewed as a combination of three of the 2--fold
symmetrical dimers bound by the weaker 3--fold symmetrical bonds. One
of the conclusions of the stochastic simulations is that predicting
the prevalent intermediaries cannot be based on energetic
considerations alone. It is possible that stable and favorable
intermediates, as determined by the static analysis, are never reached
because the necessary kinetic steps are not accessible.

\begin{figure}
  \begin{center}
    \includegraphics[width=3.25in]{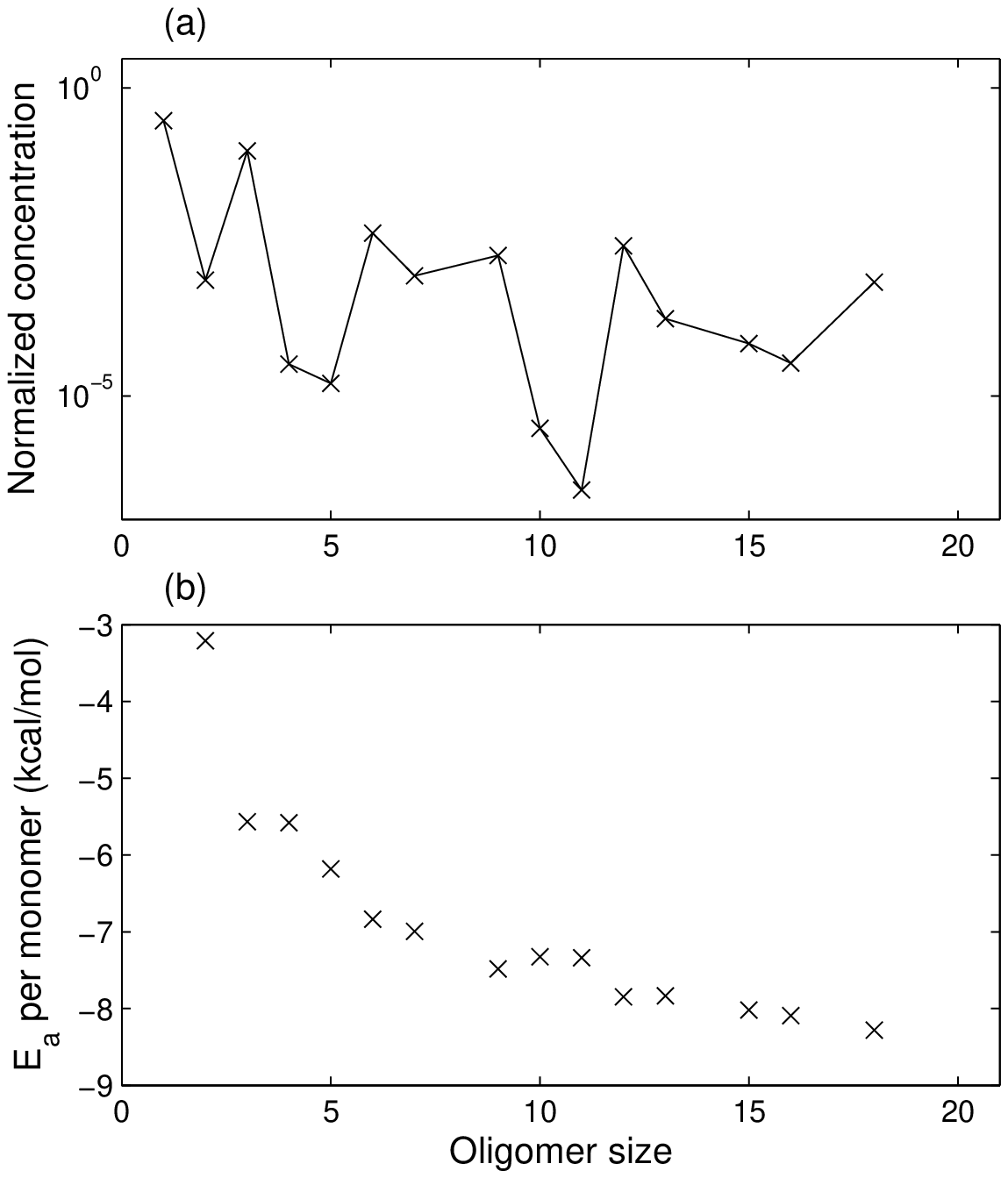}
  \caption{}
  \label{fig:4sbvConcEA}
  \end{center}
\end{figure}

\begin{figure}
  \begin{center}
   \hspace*{-.2in}  \includegraphics[width=2.1in]{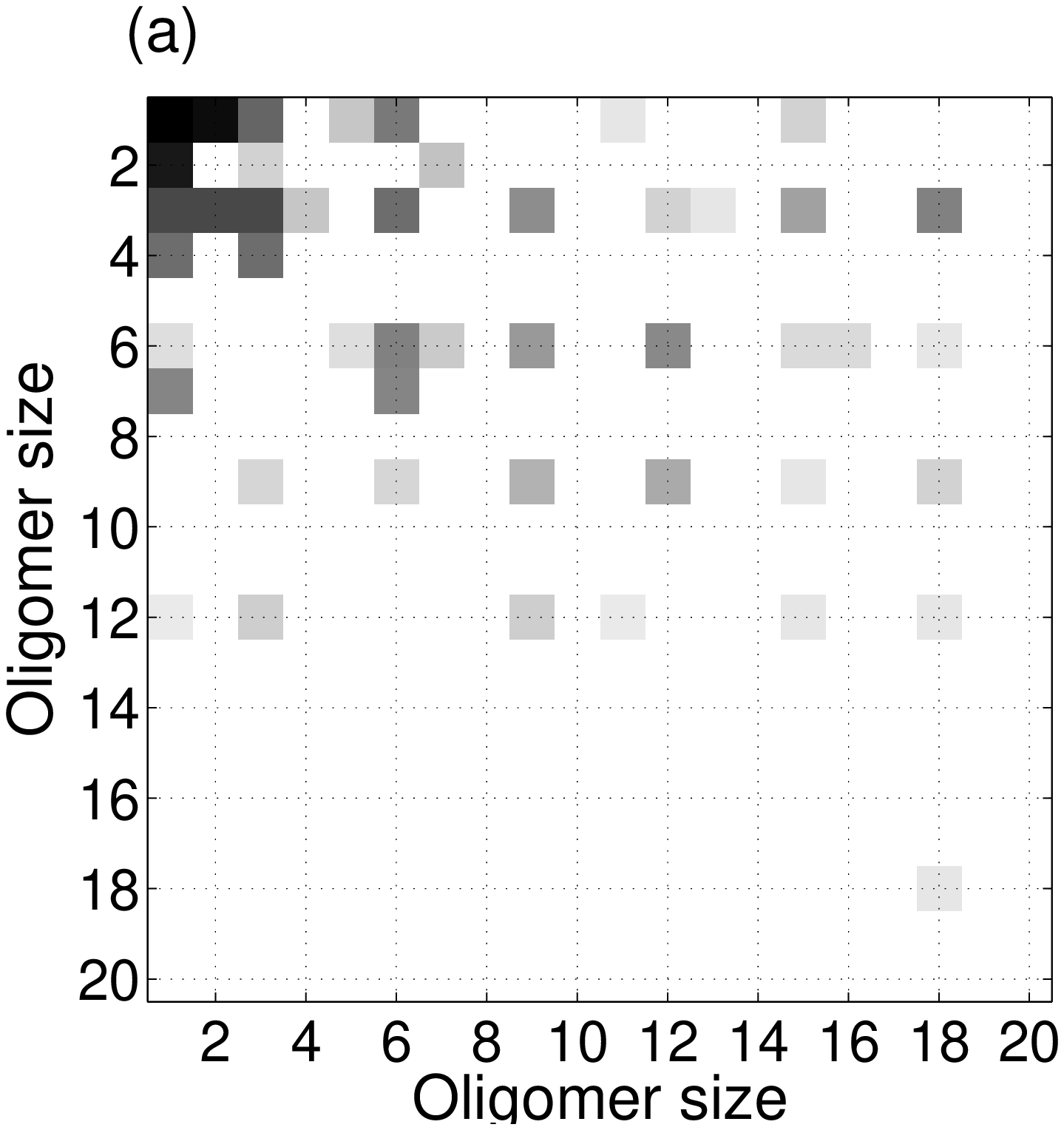}\\
    \vspace*{.2in}
    \includegraphics[width = 3.25in]{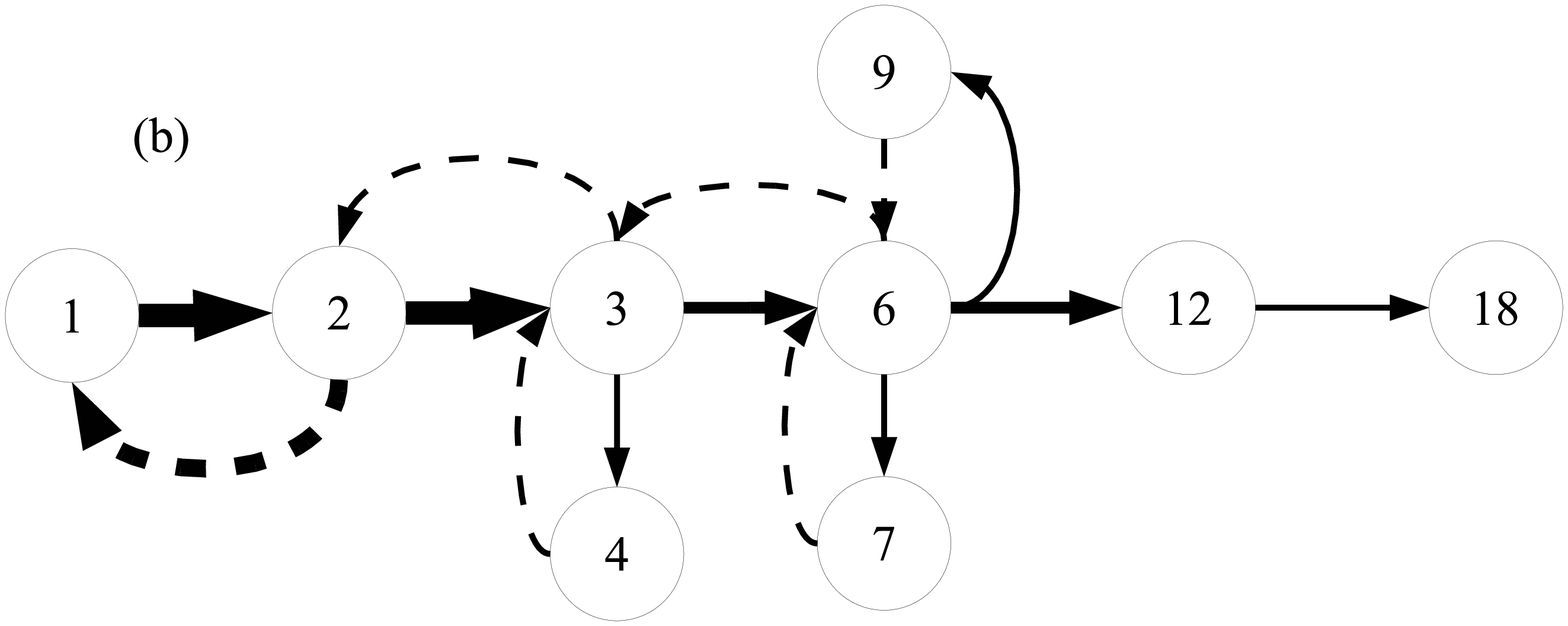}
    \caption{}
    \label{fig:4sbvHeatMap}
  \end{center}
\end{figure}

A key idea behind our method is to study how chemical properties
at the molecular level (as recorded in the protein atomic
structure) lead to differences in the assembly path. To illustrate
how our computational framework can help explore these
connections, we analyze the assembly of the $T=3$ virus SBMV
(4sbv) in direct comparison to the results obtained above for the
$T=1$ virus SPMV (1stm): Fig.~\ref{fig:4sbvConcEA} shows the
quasi-steady time-averaged concentrations and association energies
for all the oligomer sizes, while Fig.~\ref{fig:4sbvHeatMap}
presents the heat map of transitions and the relevant pathways of
assembly. The results are averaged over 1,000 runs of the
quasi-steady formation of a $>20$--mer, as before.

The average concentrations, association energies and heat maps of
4sbv reveal that trimers, hexamers, 9--mers (and, in general, all
multiples of three) have local maxima in the concentration plot
and corresponding local minima in the association energy plot.
This is also visible in the heat map as a `checkerboard pattern.'
In this case, and contrary to 1stm, trimers are the effective
units in the assembly of 4sbv, in agreement
with~\citet{Reddy:1998} and expected not only for reasons of
symmetry, but also from considering the bond energies.
Interestingly, Reddy et al.~\citep{Reddy:1998} conjecture that the
symmetric 15--mer will be the most stable oligomer. Although the
analysis with FIRST indicates that this oligomer is favorable both
in terms of association energies and of dissociation propensity,
we find no evidence of significantly higher concentration than for
other large oligomers. This could mean that although stable, this
oligomer might not be kinetically easy to access. However, it is
also possible that this is a result of our evaluation of the
energies with FIRST, as opposed to the use of energies from VIPER.

Comparing the quasi-steady concentrations of 1stm and 4sbv in
Figs.~\ref{fig:1stmConcEA} and~\ref{fig:4sbvConcEA} it is clear
that the concentration of monomers is significantly higher for
4sbv. Moreover, from the heat map (Fig.~\ref{fig:4sbvHeatMap}) it
is evident that there are more reactions involving large
oligomers. The cascading behavior is therefore less pronounced for
4sbv than for 1stm as it is less rare to find two large oligomers
present at the same time in the `solution.' This behavior stems
from the fact that the bonds in the symmetric 4sbv trimer are
significantly stronger than the bonds linking the trimers. Thus,
it is less favorable for a large 4sbv oligomer to react than it is
for a large 1stm oligomer. This is also reflected in the heat map:
since large oligomers react more slowly, there will be more
dissociation events (i.e., shaded squares below the diagonal in
Fig.~\ref{fig:4sbvHeatMap}\emph{a}) involving large oligomers for
4sbv.

\subsection*{The formation of the full capsid}

Up to now, we have focussed on the properties of the quasi-steady
state, where we assume that the supply of monomers is constant and
the cascading process of assembly leads to a stable output of
capsids. We will now examine the kinetics of formation of a full
icosahedral capsid from a finite amount of monomers.

\begin{figure}
  \begin{center}
    \includegraphics[width = 3.25in]{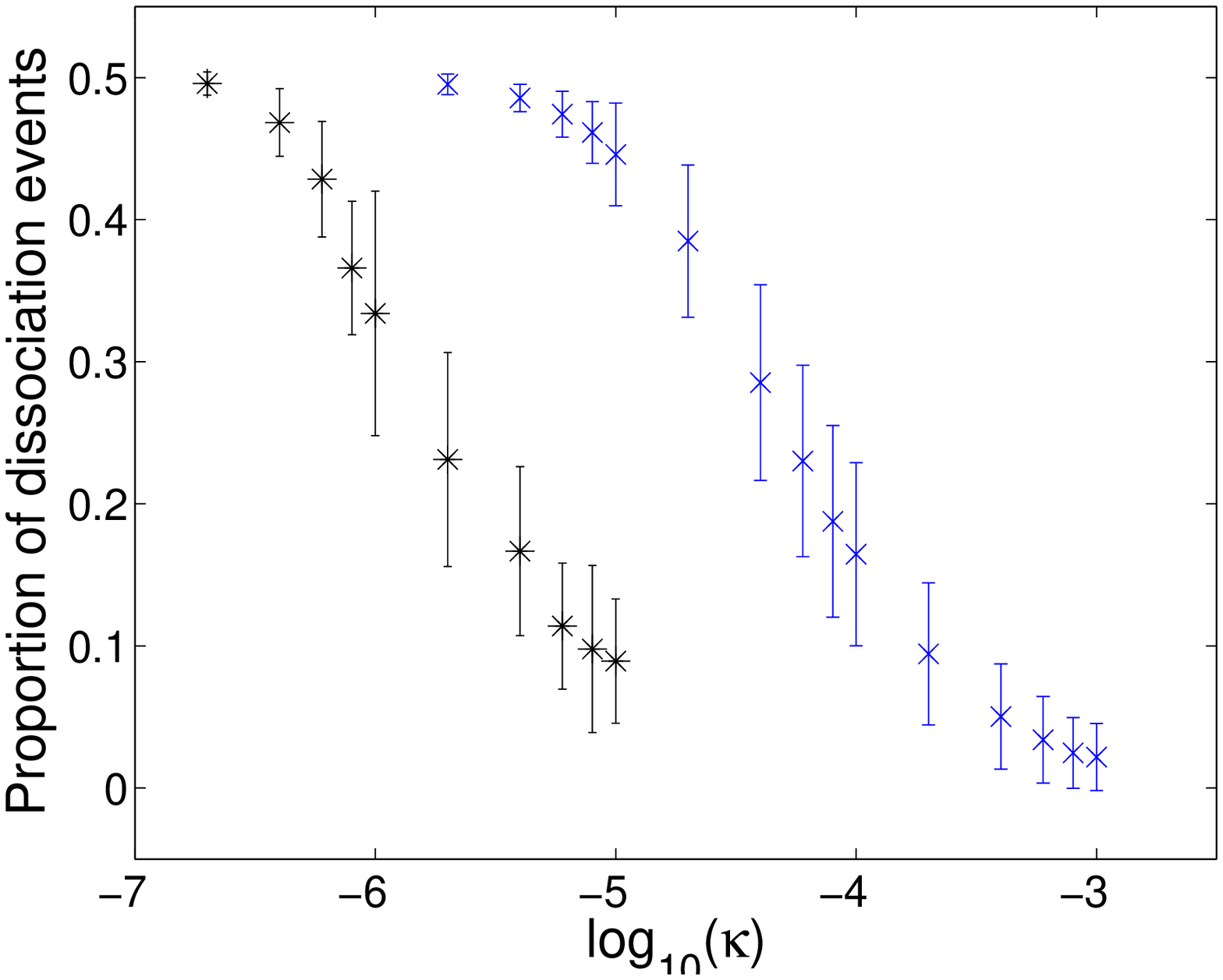}
    \caption{}
    \label{fig:1stmGeometryFactor}
  \end{center}
\end{figure}

As explained above, our dissociation calculations have an upper
limit of 20--mers, due to the use of version 3.1 of the software
FIRST. However, the cascading behavior described above for the
1stm virus implies that once large oligomers are formed it is
unlikely that they will split and thus the dissociation paths
might be ignored without much change in the observed behavior. We
have explored this idea in more detail by studying the sensitivity
of the stochastic kinetics to the form factor \geometryfactor,
which modulates the balance between the association and
dissociation pathways. Increasing \geometryfactor\ increases all
association rates, which implies that the dissociation events will
become less likely. Fig.~\ref{fig:1stmGeometryFactor} shows the
ratio between the number of dissociation events and the total
number of events in the assembly of 1stm as a function of
\geometryfactor. For low \geometryfactor\ there are almost as many
dissociation as association events, and the assembly proceeds very
slowly or not at all. In this regime, a dimer will almost
immediately be broken up once it has formed and the assembly
process is never able to get started. As \geometryfactor\
increases, there is a relatively sharp drop in the number of
dissociation events. Eventually, the number of dissociation events
becomes negligible and the assembly process proceeds almost
exclusively by association.

\begin{figure}
  \begin{center}
    \includegraphics[width = 2.1in]{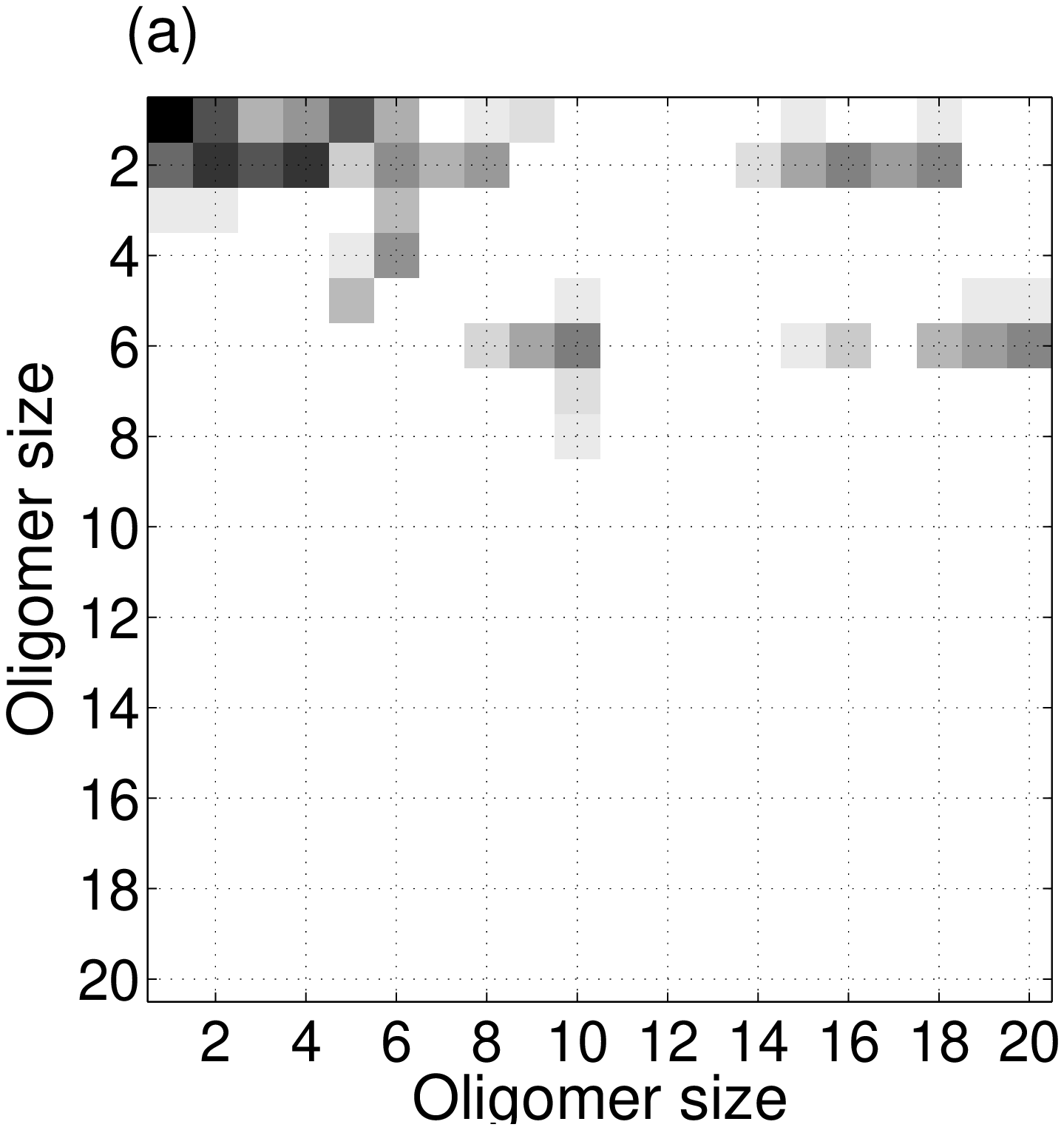}\\
    \includegraphics[width = 2.1in]{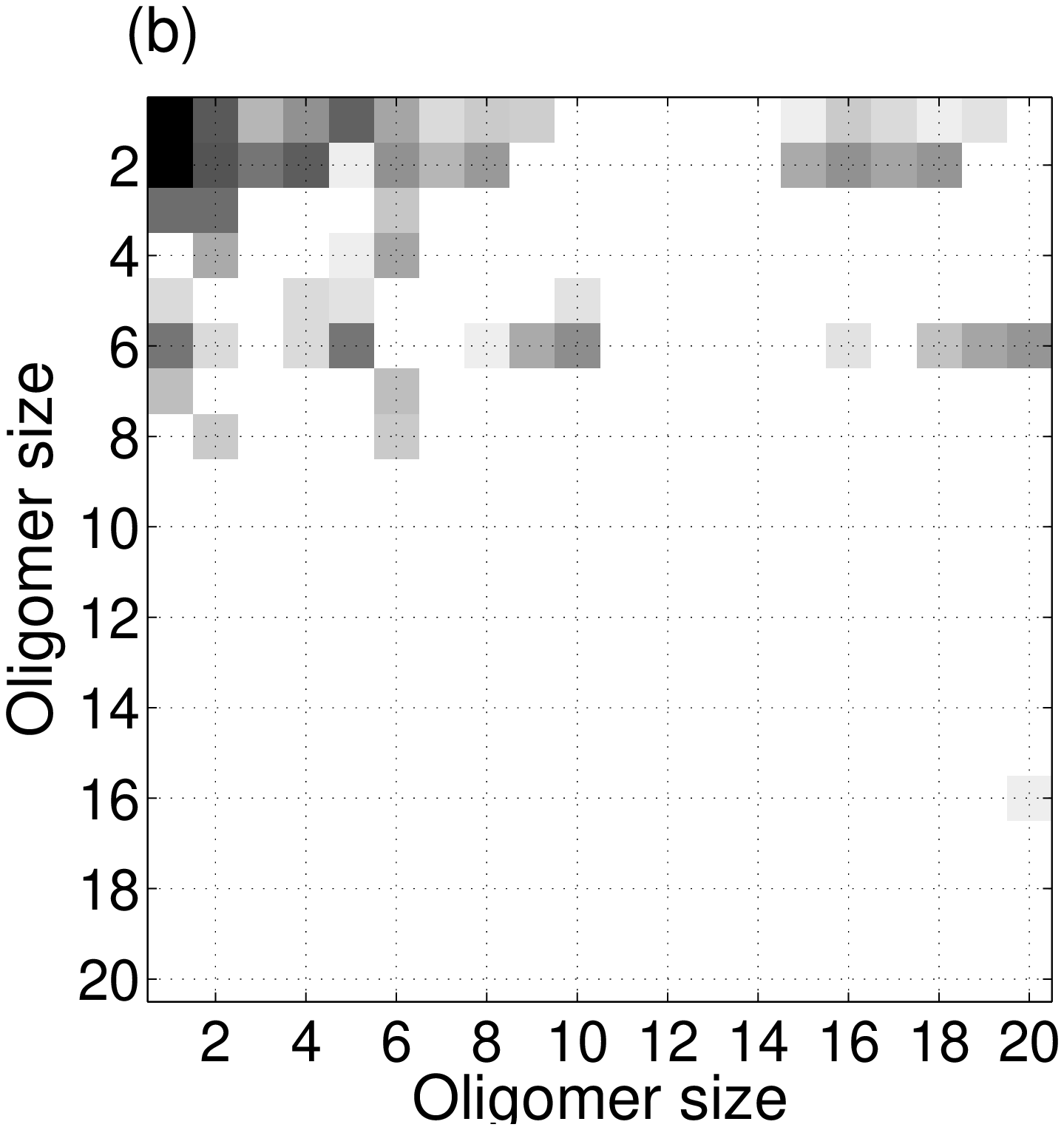}
    \caption{}
    \label{fig:1stmHeatMaps}
  \end{center}
\end{figure}

Fig.~\ref{fig:1stmHeatMaps} shows the quasi-steady concentrations
and the reaction pathways that appear in the assembly of 1stm for
different values of the form factor \geometryfactor. The key
feature of these simulations, however, is that the same types of
reactions occur for all values of \geometryfactor; that is, the
main pathways remain unchanged even if there are many dissociation
events. Under the current setup for this virus, dissociation
appears to slow the progress of aggregation by splitting small
oligomers but it does not prompt the assembly to proceed through
alternative pathways.

\begin{figure}
  \begin{center}
         \vspace{-2in}
  \centerline{\includegraphics[width = 5in]{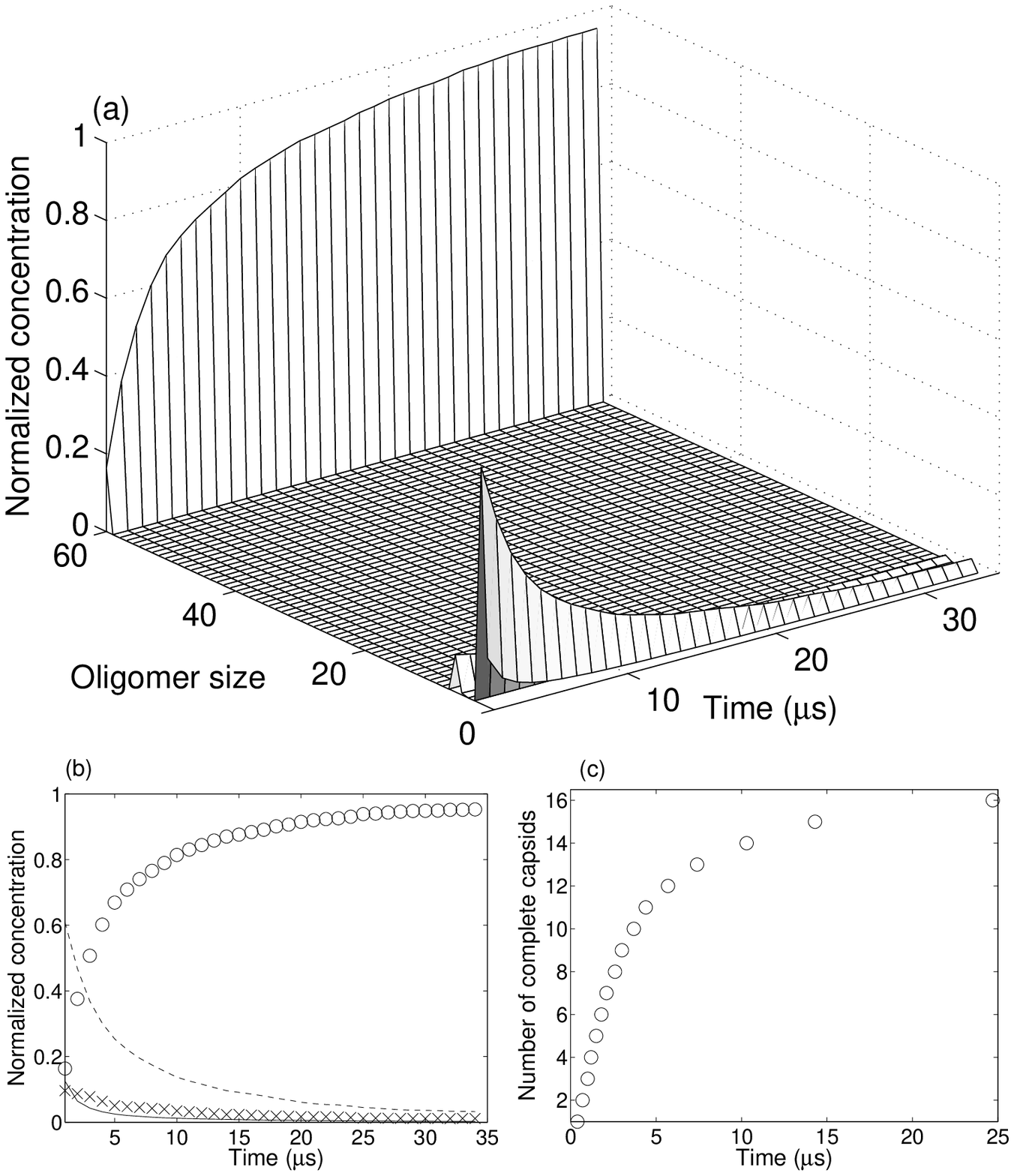}}
           \caption{}
    \label{fig:1stmCapsidCompletion}
  \end{center}
\end{figure}

A direct consequence of the particular kinetics of 1stm is that
forward (association) reactions are qualitatively similar for a
range of values of \geometryfactor . If the value of
\geometryfactor\ is relatively high, the rare dissociation events
can be neglected. We can then run simulations with association
paths alone (no longer capped by the size limit in FIRST) that
lead to the explicit formation of complete capsids. In
Fig.~\ref{fig:1stmCapsidCompletion}\emph{a,b}, we show the
concentration of the oligomers over time after the transient has
been removed for $\geometryfactor = 10^{-3}$. As expected, only
monomers, dimers, hexamers and full capsids have any significant
presence, while all other intermediate oligomers do not appear in
any persistent way. We also show in
Fig.~\ref{fig:1stmCapsidCompletion}\textit{c} that the rate of
capsid formation saturates as the concentration of monomers
decreases. The overall shape of this curve is in good agreement
with experiments and other theoretical
models~\citep{Endres:2002,Zlotnick:1999}. If we consider the
almost linear section at the outset, we can derive an approximate
capsid formation rate of $1.5 \cdot 10^{-4}$ M s$^{-1}$, which is
of the same order of magnitude as the model
by~\citet{Endres:2002}.

As a final comment, it is interesting to note that 4sbv (SBMV) can
form capsids at significantly lower \geometryfactor\ values than 1stm
(SPMV), as seen in Fig.~\ref{fig:1stmGeometryFactor}. This conclusion
cannot be drawn easily from the association energies alone: the most
favorable 1stm dimer and 4sbv dimer have association energies of $-48$
and $-38$ kcal/mol, respectively. However, when a 4sbv dimer is
formed, a favorable reaction to form a symmetric trimer tends to
follow immediately. On the other hand, despite their higher binding
energy, the 1stm dimers have no such favorable aggregation pathway to
form a stable large oligomer.

\section*{Discussion and conclusions}

\subsection*{Understanding the quasi-steady solution as a Markov process}

Our Gillespie simulation of the kinetics of the network has shown
that although the full assembly tree
(Fig.~\ref{fig:1stmAssemblyTree}) is extremely complex, only a few
pathways are crucial for the assembly. In other words, our
extended Gillespie algorithm provides us with a stochastic
sampling of the reaction network, unknown \emph{a priori}, which
leads to an estimate of the transition probabilities in the
system.

We can use the estimated transition matrix (represented in
Figs.~\ref{fig:1stmReactionCartoon}\emph{a}
and~\ref{fig:4sbvHeatMap}\emph{a} as heat maps), to investigate the
description of the reaction network as a non-homogeneous Markov
process. To check the consistency of the quasi-steady solution
obtained numerically in Figs.~\ref{fig:1stmConcEA}
and~\ref{fig:4sbvConcEA}, we apply the results of~\citet{Darroch:1965}
for quasi-stationary Markov processes taking the stoichiometry into
consideration~\citep{Antia:1985}.  The system is only
\textit{quasi}-stationary because ultimately there is an absorbing
state where all monomers are part of completed capsids. However, in
the transient state the quasi-stationary distribution (QSD) can be
calculated as the fixed point $\pi^*$ of the following equation:
\begin{equation}
  \label{eq:QSD}
  b(\pi) = \frac{\pi^TQ^{-1}}{\pi^TQ^{-1}e} \,  \Rightarrow
  \quad{\pi^*}^T=b(\pi^*),
\end{equation}
where $\pi$ is the distribution of concentrations, $e$ is the vector
of ones, and $Q$ is the transition rate matrix as derived from the
simulations~\citep{Darroch:1967}. There are two interpretations of the
QSD~\citep{Darroch:1965}: it can be viewed as a conditional stationary
distribution (i.e., the stationary distribution provided that the
Markov process is in the transient), or as the expected time spent in
each state divided by the total time to
absorption. Fig.~\ref{fig:1stmConcEA}\textit{a} shows that the QSD
$\pi^*$ is close to the average empirical distribution $\pi$ from the
simulations. The transition rate matrix derived using the stochastic
sampling is therefore consistent with the observed quasi-steady
distribution under the assumption of a non-homogeneous semi-Markov
process. This description also provides us insight into why some
oligomers have a Gaussian-like distribution while others present
Poisson-like features. In a system where the total number of monomers
is fixed, the (quasi) stationary distribution will be
multinomial~\citep{Gadgil:2005}. In the limit of large $N$ and small
$\pi_i$, the distribution of oligomer $i$ can be approximated by a
Poisson distribution. For large $N$ and intermediate $\pi_i$, the
Gaussian distribution is a good approximation.

\subsection*{Explaining the probabilistic features of the cascading behavior}

In the cascading process, a few oligomers are relatively
long-lived while all other oligomers survive for only very short
times before reacting. The existence of Gaussian-like and
Poisson-like distributions is related to this cascading process
and can be understood through a simplified kinetic model. Consider
a toy model of the early stages of aggregation of 1stm consisting
of three oligomers ($M_2$, $M_4$ and $M_6$), which can be thought
of as analogues of the dimers, tetramers and hexamers,
respectively:
\begin{eqnarray} \label{eq:simple_reaction}
  \emptyset & \xrightarrow{k_H} & M_2  \nonumber \\
  2M_2 & \xrightarrow{k_L} & M_4 \nonumber \\
  M_2 + M_4 & \xrightarrow{k_H} & M_6 \nonumber \\
  M_6 & \xrightarrow{k_L} & \emptyset.
\end{eqnarray}
The first and last reactions correspond to creation from a source
and decay to a sink and there are two reaction rates $k_H \gg
k_L$. We simulate this system using the Gillespie algorithm. The
resulting stationary distributions, shown in
Fig.~\ref{fig:1stmToySystem}, present similar characteristics to
those discussed in the 1stm assembly process (see
Fig.~\ref{fig:1stmPoissonGaussian}). This can be understood as
follows: the creation rate of the `dimers' $M_2$ is much higher
than the rate at which they are consumed, leading to a Gaussian
steady state, as predicted by the linear noise approximation of
van Kampen~\citep{Hayot:2004,vanKampen:1992}. On the other hand,
the `tetramers' $M_4$ have a low creation rate and there are
always `dimers' available with which they can react at a high
rate. This leads to a Poisson-like distribution for the
`tetramers.' Finally, `tetramers' disappear quickly to create
`hexamers' $M_6$, which decay at a very low rate and thus have a
Gaussian-like distribution.

\subsection*{Summary and future work}

This paper presents a modular framework for the study of the
stochastic kinetics of viral capsid assembly. The calculations are
based on structural crystallographic protein data and use rigidity
analysis to produce a reduced mechanical description of the
protein oligomers. Rates for association and dissociation
reactions based on the protein descriptions are then used within
an extended Gillespie algorithm to explore the kinetics of capsid
assembly.

Because of its biophysical motivation, our model has relatively
few parameters and most of them are directly related to physical
variables: temperature, concentration, diffusion coefficients,
protein radius, covalent bond energies, and bond lengths. We have
checked the dependence of our simulations on these physical
variables. For instance, if the temperature is increased,
dissociation events will become more likely and the overall rate
of assembly will drop. In addition, the relative difference in
association energies between oligomers decreases. This means that
the population of oligomers will become more varied and more
reaction pathways will become important; that is, as the
temperature increases, the assembly tree will be explored more
homogeneously. Similarly, lowering the concentration decreases the
association rate. If the concentration is too low, the
dissociation events become prevalent and there will be no
assembly. However, the characteristics of the assembly pathways
are unchanged by concentration.

There are three additional parameters (\ecut, \geometryfactor\ and
\waterfactor) that have physical meaning and motivation, but are not
easily related to a single physical variable. We have discussed in
depth the effect of the form factor \geometryfactor\ in the preceding
sections. In addition, we have checked that the results of our
analysis do not depend qualitatively on the cut-off energy \ecut\ or
the water shielding constant \waterfactor.  Increasing the cut-off
energy for hydrogen bonds in FIRST reduces the number of hydrogen
bonds in the system. This produces the same overall effect as
increasing the temperature since all energies in the system are
lowered. It also leads to floppier proteins with a higher eigenratio
in Eq.~\ref{eq:multidimensionEscapeRate} and thus more dissociation
events. Increasing the water shielding \waterfactor\ means stronger
hydrogen bonds, which is equivalent to lowering the temperature. The
assembly will thus proceed along low energy pathways, with a small
variety of oligomers and a reduced number of dissociation events. This
discussion indicates that changes in both \ecut\ and \waterfactor\ can
be qualitatively understood as an effective change of `temperature.'
Note however that the effect of the form factor \geometryfactor\ is
different. Physically, the increase of \geometryfactor\ is equivalent
to lowering the barrier for two oligomers to form a larger oligomer
with no influence on the dissociation process. Therefore, the
likelihood of the dissociation events is reduced and the assembly is
sped up.

A key feature of the proposed framework is that it is both modular and
extensible, i.e., the algorithms that make up the different components
of the model can be exchanged seamlessly at different levels. A number
of refinements to the model should be pursued to improve the
oversimplifications of this initial work. Indeed, the bond energies
could be calculated more precisely using more detailed
potentials. This can have far-reaching implications for the pathways
and intermediates deduced from the simulations and a variety of energy
calculations should be explored carefully when dealing with specific
viruses~\citep{Zhang:2006}.  A key ingredient of the protein model is
the derivation of a reduced representation from the full PDB data. In
this work, we have used FIRST for protein partitioning as a conceptual
tool based on ideas from graph rigidity. However, one could use
methods based on normal modes (full atom, elastic or gaussian models)
or principal component analysis to obtain coarse-grained
representations of proteins. Another important set of refinements
should concentrate on the description of the association process. In
particular, a more sophisticated model of the protein docking,
including its entropic aspects, would be necessary to improve the
physical realism of the form factor \geometryfactor . Moreover, it
would be important to refine the association rates to parallel more
closely the kinetics of chemical assembly. The dissociation model
itself could also be improved by taking explicitly into account
entropic features and incorporating the geometric content of the graph
when computing the eigenfrequencies. Finally, it would be important
(although non-trivial) to extend our model to allow for
non-icosahedral symmetries and for malformed
capsids~\citep{Berger:1994, Rapaport:2004}.

In summary, our work introduces a description of viral capsid
formation as a stochastic assembly of protein oligomers. An
important aspect is that our framework is data-driven, starting
from molecular detail, and exhibits different assembly behaviors
for different viruses, as exemplified by the results for 1stm and
4sbv presented here. Importantly, no assumptions are made about
specific intermediates through which the assembly has to
proceed---all such phenomena emerge from the data. Our methodology
bridges the gap between the static and dynamic models of viral
assembly by using a stochastic sampling algorithm to investigate
the assembly pathways.  The sampling is done using an extended
version of the Gillespie algorithm which is derived from
fundamental physical principles. This enables a mesoscopic
simulation of the kinetics which is less computationally intensive
than a microscopic MD simulation. Alternatively, this algorithm
provides a physically based sampling of the assembly tree, as
opposed to computationally intractable combinatorial optimization
techniques~\citep{Reddy:1998, Horton:1992}. We are currently in
the process of extending and refining our framework in several of
the above directions as we pursue a general exploration of other
icosahedral viruses in different families.

\vspace*{.5cm}

\begin{small}
We should like to thank D.J.\ Jacobs, B.M.\ Hespenheide and M.F.\
Thorpe for permission to use the FIRST software, and for their
responsiveness to our queries. We are also grateful to Gert Vriend
for access to the WHAT IF software. We gratefully acknowledge
helpful discussions with Peter H\"anggi, Vijay Reddy, Christina
Kyriakidou and Sophie Scialom. Research funded by EPSRC, the Royal
Society, and the Institute of Biomedical Engineering at Imperial
College London.
\end{small}

\bibliography{./virus}

\begin{thebibliography}{51}
\expandafter\ifx\csname natexlab\endcsname\relax\def\natexlab#1{#1}\fi

\bibitem[{Alberts et~al.(2002)Alberts, Johnson, Lewis, Raff, Roberts, and
  Walter}]{Alberts:2002}
Alberts, B., A.~Johnson, J.~Lewis, M.~Raff, K.~Roberts, and P.~Walter. 2002.
\newblock Molecular Biology of the Cell.
\newblock 4th edition. Garland Science.

\bibitem[{Caspar and Klug(1962)}]{Caspar:1962}
Caspar, D. L.~D., and A.~Klug. 1962.
\newblock Physical principles in the construction of regular viruses.
\newblock \emph{In} Cold Spring Harb. Symp. Quant. Biol., volume~27. 1--24.

\bibitem[{Zlotnick(2004)}]{Zlotnick:2004}
Zlotnick, A. 2004.
\newblock Viruses and the physics of soft condensed matter.
\newblock \emph{Proceedings of the National Academy of Sciences}
  101:15549--15550.

\bibitem[{Zlotnick(2003)}]{Zlotnick:2003}
Zlotnick, A. 2003.
\newblock Are weak protein-protein interactions the general rule in capsid
  assembly.
\newblock \emph{Virology} 315:269--274.

\bibitem[{McPherson(2005)}]{McPherson:2005}
McPherson, A. 2005.
\newblock Micelle formation and crystallization as paradigms for virus
  assembly.
\newblock \emph{BioEssays} 27:447--458.

\bibitem[{Schwartz et~al.(1998)Schwartz, Shor, Jr, and Berger}]{Schwartz:1998}
Schwartz, R., P.~W. Shor, P.~E.~P. Jr, and B.~Berger. 1998.
\newblock Local rules simulation of the kinetics of virus capsid self-assembly.
\newblock \emph{Biophysical Journal} 75:2626--2636.

\bibitem[{Larson and McPherson(2001)}]{Larson:2001}
Larson, S.~B., and A.~McPherson. 2001.
\newblock Satellite {T}obacco {M}osaic {V}irus {RNA}: structure and
  implications for assembly.
\newblock \emph{Current Opinion in Structural Biology} 11:59--65.

\bibitem[{Fox et~al.(1998)Fox, Wang, Speir, Olson, Johnson, Baker, and
  Young}]{Fox:1998}
Fox, J.~M., G.~Wang, J.~A. Speir, N.~H. Olson, J.~E. Johnson, T.~S. Baker, and
  M.~J. Young. 1998.
\newblock Comparison of the native {CCMV} virion with in vitro assembled {CCMV}
  virions by cryoelectron microscopy and image reconstruction.
\newblock \emph{Virology} 244:212--218.

\bibitem[{Baker et~al.(1999)Baker, Olson, and Fuller}]{Baker:1999}
Baker, T.~S., N.~H. Olson, and S.~D. Fuller. 1999.
\newblock Adding the third dimension to virus life cycles: Three-dimensional
  reconstruction of icosahedral viruses from cryo-electron micrographs.
\newblock \emph{Microbiology and Molecular Biology Reviews} 63:862--922.

\bibitem[{Berger et~al.(1994)Berger, Shor, Tucker-Kellogg, and
  King}]{Berger:1994}
Berger, B., P.~W. Shor, L.~Tucker-Kellogg, and J.~King. 1994.
\newblock Local rule-based theory of virus shell assembly.
\newblock \emph{Proceedings of the National Academy of Sciences} 91:7732--7736.

\bibitem[{Rapaport(2004)}]{Rapaport:2004}
Rapaport, D. 2004.
\newblock Self-assembly of polyhedral shells: A molecular dynamics study.
\newblock \emph{Physical Review E} 70:051905.

\bibitem[{Horton and Lewis(1992)}]{Horton:1992}
Horton, N., and M.~Lewis. 1992.
\newblock Calculation of the free energy of association for protein complexes.
\newblock \emph{Protein Science} 1:169--181.

\bibitem[{Reddy et~al.(1998)Reddy, Giesing, Morton, Kumar, Post, III, and
  Johnson}]{Reddy:1998}
Reddy, V.~S., H.~A. Giesing, R.~T. Morton, A.~Kumar, C.~B. Post, C.~L.~B. III,
  and J.~E. Johnson. 1998.
\newblock Energetics of quasiequivalence: Computational analysis of
  protein-protein interactions in icosahedral viruses.
\newblock \emph{Biophysical Journal} 74:546--558.

\bibitem[{Sitharam and Agbandje-McKenna(2006)}]{Sitharam:2004}
Sitharam, M., and M.~Agbandje-McKenna. 2006.
\newblock Modeling virus self-assembly pathways: Avoiding dynamics using
  geometric constraint decomposition.
\newblock \emph{Journal of Computational Biology} Accepted for publication.

\bibitem[{Hespenheide et~al.(2004)Hespenheide, Jacobs, and
  Thorpe}]{Hespenheide:2004}
Hespenheide, B.~M., D.~J. Jacobs, and M.~F. Thorpe. 2004.
\newblock Structural rigidity in the capsid assembly of {C}owpea {C}hlorotic
  {M}ottle {V}irus.
\newblock \emph{Journal of Physics: Condensed Matter} 16:5055--5064.

\bibitem[{Endres and Zlotnick(2002)}]{Endres:2002}
Endres, D., and A.~Zlotnick. 2002.
\newblock Model-based analysis of assembly kinetics for virus capsids or other
  spherical polymers.
\newblock \emph{Biophysical Journal} 83:1217--1230.

\bibitem[{Zlotnick et~al.(1999)Zlotnick, Johnson, Wingfield, Stahl, and
  Endres}]{Zlotnick:1999}
Zlotnick, A., J.~M. Johnson, P.~W. Wingfield, S.~J. Stahl, and D.~Endres. 1999.
\newblock A theoretical model successfully identifies features of {H}epatitis
  {B} virus capsid assembly.
\newblock \emph{Biochemistry} 38:14644--14652.

\bibitem[{Ceres and Zlotnick(2002)}]{Ceres:2002}
Ceres, P., and A.~Zlotnick. 2002.
\newblock Weak protein-protein interactions are sufficient to drive assembly of
  {H}epatitis {B} virus capsids.
\newblock \emph{Biochemistry} 41:11525--11531.

\bibitem[{Kegel and van~der Schoot(2004)}]{Kegel:2004}
Kegel, W.~K., and P.~van~der Schoot. 2004.
\newblock Competing hydrophobic and screened-coulomb interactions in
  {H}epatitis {B} virus capsid assembly.
\newblock \emph{Biophysical Journal} 86:3905--3913.

\bibitem[{Lidmar et~al.(2003)Lidmar, Mirny, and Nelson}]{Lidmar:2003}
Lidmar, J., L.~Mirny, and D.~R. Nelson. 2003.
\newblock Virus shapes and buckling transitions in spherical shells.
\newblock \emph{Physical Review E} 68:0306741.

\bibitem[{Marzec and Day(1993)}]{Marzec:1993}
Marzec, C., and L.~Day. 1993.
\newblock Pattern formation in icosahedral virus capsids: the {P}apova viruses
  and {N}udaurelia {C}apensis {B}eta virus.
\newblock \emph{Biophysical Journal} 65:2559--2577.

\bibitem[{Tarnai et~al.(1995)Tarnai, Gaspar, and Szalai}]{Tarnai:1995}
Tarnai, T., Z.~Gaspar, and L.~Szalai. 1995.
\newblock Pentagon packing models for "all-pentamer" virus structures.
\newblock \emph{Biophysical Journal} 69:612--618.

\bibitem[{Bruinsma et~al.(2003)Bruinsma, Gelbart, Reguera, Rudnick, and
  Zandi}]{Bruinsma:2003}
Bruinsma, R.~F., W.~M. Gelbart, D.~Reguera, J.~Rudnick, and R.~Zandi. 2003.
\newblock Viral self-assembly as a thermodynamic process.
\newblock \emph{Physical Review Letters} 90:248101.

\bibitem[{Zandi et~al.(2004)Zandi, Reguera, Bruinsma, Gelbart, and
  Rudnick}]{Zandi:2004}
Zandi, R., D.~Reguera, R.~F. Bruinsma, W.~M. Gelbart, and J.~Rudnick. 2004.
\newblock Origin of icosahedral symmetry in viruses.
\newblock \emph{Proceedings of the National Academy of Sciences}
  101:15556--15560.

\bibitem[{Endres et~al.(2005)Endres, Miyahara, Moisant, and
  Zlotnick}]{Endres:2005}
Endres, D., M.~Miyahara, P.~Moisant, and A.~Zlotnick. 2005.
\newblock A reaction landscape identifies the intermediates critical for
  self-assembly of virus capsids and other polyhedral structures.
\newblock \emph{Protein Science} 14:1518--1525.

\bibitem[{Reddy et~al.(2001)Reddy, Natarajan, Okerberg, Li, Damodaran, Morton,
  III, and Johnson}]{Reddy:2001}
Reddy, V., P.~Natarajan, B.~Okerberg, K.~Li, K.~Damodaran, R.~Morton, C.~B.
  III, and J.~Johnson. 2001.
\newblock {VI}rus {P}article {E}xplore{R} ({VIPER}), a website for virus capsid
  structures and their computational analyses.
\newblock \emph{Journal of Virology} 75:11943--11947.

\bibitem[{Jacobs et~al.(2001)Jacobs, Rader, Kuhn, and Thorpe}]{Jacobs:2001}
Jacobs, D.~J., A.~Rader, L.~A. Kuhn, and M.~Thorpe. 2001.
\newblock Protein flexibility predictions using graph theory.
\newblock \emph{Proteins: Structure, Function and Genetics} 44:150--165.

\bibitem[{Gillespie(1977)}]{Gillespie:1977}
Gillespie, D.~T. 1977.
\newblock Exact stochastic simulation of coupled chemical reactions.
\newblock \emph{Journal of Physical Chemistry} 81:2340--2361.

\bibitem[{Vriend(1990)}]{Vriend:1990}
Vriend, G. 1990.
\newblock {WHAT IF}: A molecular modeling and drug design program.
\newblock \emph{J Mol Graph} 8:52--56.

\bibitem[{Jacobs(1998)}]{Jacobs:1998}
Jacobs, D.~J. 1998.
\newblock Generic rigidity in three-dimensional bond-bending networks.
\newblock \emph{Journal of Physics A} 31:6653--6668.

\bibitem[{Gillespie(1976)}]{Gillespie:1976}
Gillespie, D.~T. 1976.
\newblock A general method for numerically simulating the stochastic time
  evolution of coupled chemical reactions.
\newblock \emph{Journal of Computational Physics} 22:403--434.

\bibitem[{van Kampen(1992)}]{vanKampen:1992}
van Kampen, N.~G. 1992.
\newblock Stochastic processes in physics and chemistry.
\newblock 2nd edition. Elsevier.

\bibitem[{Turner et~al.(2004)Turner, Schnell, and Burrage}]{Turner:2004}
Turner, T., S.~Schnell, and K.~Burrage. 2004.
\newblock Stochastic approaches for modelling in vivo reactions.
\newblock \emph{Computational Biology and Chemistry} 28:165--178.

\bibitem[{Selzer and Schreiber(2001)}]{Selzer:2001}
Selzer, T., and G.~Schreiber. 2001.
\newblock New insights into the mechanism of protein-protein association.
\newblock \emph{Proteins: Structure, Function and Genetics} 45:190--198.

\bibitem[{Janin(1997)}]{Janin:1997}
Janin, J. 1997.
\newblock The kinetics of protein-protein recognition.
\newblock \emph{Proteins: Structure, Function and Genetics} 28:153--161.

\bibitem[{Stundzia and Lumsden(1996)}]{Stundzia:1996}
Stundzia, A.~B., and C.~J. Lumsden. 1996.
\newblock Stochastic simulation of coupled reaction-diffusion processes.
\newblock \emph{Journal of Computational Physics} 127:196--207.

\bibitem[{Bernstein(2005)}]{Bernstein:2005}
Bernstein, D. 2005.
\newblock Simulating mesoscopic reaction-diffusion systems using the
  {G}illespie algorithm.
\newblock \emph{Physical Review E} 71:041103.

\bibitem[{Drake(1972)}]{Drake:1972}
Drake, R.~L. 1972.
\newblock Aerosol physics and chemistry, volume~3, chapter~2.
\newblock 1st edition. Pergamon Press Inc, 201--377.

\bibitem[{Camacho et~al.(1999)Camacho, Weng, Vajda, and DeLisi}]{Camacho:1999}
Camacho, C.~J., Z.~Weng, S.~Vajda, and C.~DeLisi. 1999.
\newblock Free energy landscapes of encounter complexes in protein-protein
  association.
\newblock \emph{Biophysical Journal} 76:1166--1178.

\bibitem[{Northrup and Erickson(1992)}]{Northrup:1992}
Northrup, S., and H.~Erickson. 1992.
\newblock Kinetics of protein-protein association explained by {B}rownian
  dynamics computer simulation.
\newblock \emph{Proceedings of the National Academy of Sciences} 89:3338--3342.

\bibitem[{Sheu et~al.(2003)Sheu, Yang, H.~L, and Schlag}]{Sheu:2003}
Sheu, S.-Y., D.-Y. Yang, S.~H.~L, and E.~W. Schlag. 2003.
\newblock Energetics of hydrogen bonds in peptides.
\newblock \emph{Proceedings of the National Academy of Sciences}
  100:12683--12687.

\bibitem[{Brooks et~al.(1983)Brooks, Bruccoleri, Olafson, States, Swaminathan,
  and Karplus}]{Brooks:1983}
Brooks, B.~R., R.~E. Bruccoleri, B.~D. Olafson, D.~J. States, S.~Swaminathan,
  and M.~Karplus. 1983.
\newblock {CHARMM}: A program for macromolecular energy, minimization, and
  dynamics calculations.
\newblock \emph{Journal of Computational Chemistry} 4:187--217.

\bibitem[{H\"{a}nggi et~al.(1990)H\"{a}nggi, Talkner, and
  Borkovec}]{Hanggi:1990}
H\"{a}nggi, P., P.~Talkner, and M.~Borkovec. 1990.
\newblock Reaction-rate theory: fifty years after {K}ramers.
\newblock \emph{Reviews of Modern Physics} 62:251--342.

\bibitem[{Savithri and Erickson(1983)}]{Savithri:1983}
Savithri, H., and J.~Erickson. 1983.
\newblock The self-assembly of the {C}owpea strain of the {S}outhern {B}ean
  {M}osaic {V}irus.
\newblock \emph{Virology} 126:328--335.

\bibitem[{Darroch and Seneta(1965)}]{Darroch:1965}
Darroch, J.~N., and E.~Seneta. 1965.
\newblock On quasi-stationary distributions in absorbing discrete-time finite
  {M}arkov chains.
\newblock \emph{Journal of Applied Probability} 2:88--100.

\bibitem[{Antia and Lee(1985)}]{Antia:1985}
Antia, F.~D., and S.~Lee. 1985.
\newblock The effect of stoichiometry on {M}arkov chain models for chemical
  reaction kinetics.
\newblock \emph{Chemical Engineering Science} 40:1969--1971.

\bibitem[{Darroch and Seneta(1967)}]{Darroch:1967}
Darroch, J.~N., and E.~Seneta. 1967.
\newblock On quasi-stationary distributions in absorbing continuous-time finite
  {M}arkov chains.
\newblock \emph{Journal of Applied Probability} 4:192--196.

\bibitem[{Gadgil et~al.(2005)Gadgil, Lee, and Othmer}]{Gadgil:2005}
Gadgil, C., C.-H. Lee, and H.~G. Othmer. 2005.
\newblock A stochastic analysis of first-order reaction networks.
\newblock \emph{Bulletin of Mathematical Biology} 67:901--946.

\bibitem[{Hayot and Jayaprakash(2004)}]{Hayot:2004}
Hayot, F., and C.~Jayaprakash. 2004.
\newblock The linear noise approximation for molecular fluctuations within the
  cell.
\newblock \emph{Physical Biology} 1:205--210.

\bibitem[{Zhang and Schwartz(2006)}]{Zhang:2006}
Zhang, T., and R.~Schwartz. 2006.
\newblock Simulation study of the contribution of oligomer/oligomer binding to
  capsid assembly kinetics.
\newblock \emph{Biophysical Journal} 90:57--64.

\bibitem[{Sayle and Milner-White(1995)}]{Sayle:1995}
Sayle, R., and E.~Milner-White. 1995.
\newblock {RASMOL}: biomolecular graphics for all.
\newblock \emph{Trends in Biochemical Sciences} 20:374--376.

\end{thebibliography}

\clearpage
\section*{Tables}

\begin{table}[!htbp]
  \centering
  \begin{tabular}{cccc}
    \hline
    Interface & Symmetry & $E^{(a)}_{ij}$ VIPER & $E^{(a)}_{ij}$ FIRST \\
     & & (kcal/mol) & (kcal/mol) \\
    \hline
    1--6 & Quasi 5--fold & $-21.0$ & $-9.0$ \\
    1--38 & Quasi 5--fold & $-21.0$ & $-9.0$ \\
    1--37 & Quasi 2--fold & $-29.0$ & $-48.7$ \\
    1--2 & Quasi 3--fold & $-33.0$ & $-24.7$ \\
    1--3 & Quasi 3--fold & $-33.0$ & $-24.7$ \\
    \hline
  \end{tabular}
\caption{\textbf{Association energies from FIRST and VIPER.} A
comparison of the association energies for the Satellite Panicum
Mosaic Virus (SPMV, PDB code 1stm) computed using
VIPER~\citep{Reddy:2001} and with FIRST with \ecut $= - 0.7$
kcal/mol. The interfaces are shown in
Fig.~\protect{\ref{fig:1stmInterfaces}}\textit{a}. Note that in
the simulations these energies are multiplied by the water
shielding factor \waterfactor = 0.17 to account for protein
hydration.} \label{tab:interfaceEnergies}
\end{table}

\begin{table}[!htbp]
  \centering
  \begin{tabular}{lrl}
    \hline
    Parameter & Value & Units \\
    \hline
    Temperature, T & $300$ & K \\
    Initial monomer concentration, C & $5$ & $\mu$M \\
    Monomer diffusion coefficient, $D_1$ & $0.1$ & nm$^2$/s \\
    Monomer radius, $r_1$ & $1$ & nm \\
    Covalent bond strength & $-74$ & kcal/mol \\
    Covalent bond length & $1.5$ & \AA \\
    Hydrogen bond length & $3$ & \AA \\
    \\
    H-bond effective strength, \waterfactor & $0.17$ & \quad --- \\
    FIRST cut-off energy, \ecut & $-0.7$ & kcal/mol \\
    \hline
  \end{tabular}
  \caption{\textbf{Parameter values for the simulations.}
  Besides those in the table, there is an additional
  parameter in the simulations: the form factor \geometryfactor\,
  which is initially chosen to be $10^{-4}$ for 1stm and $2 \cdot 10^{-5}$
  for 4sbv. The dependence of the results on
  \geometryfactor\ is shown in Figs.~\ref{fig:1stmGeometryFactor}
  and~\ref{fig:1stmHeatMaps}.}
  \label{tab:parameterValues}
\end{table}

\clearpage

\section*{Figure Legends}

\subsubsection*{Figure~\ref{fig:1stmInterfaces}}
\textbf{The icosahedral geometry of a $T = 1$ capsid.}
(\textit{a}) There are 60 symmetrically equivalent lattice
positions, each one occupied by an asymmetric protein. For the
1stm virus, the protein positioned at 1 has bonds with those
positioned at 2, 3, 6, 38 and 37 (but not with 4), with energies
as shown in Table~\protect{\ref{tab:interfaceEnergies}}.
(\textit{b}) A flattened view of the icosahedron above.

\subsubsection*{Figure~\ref{fig:1stmFlexibility}}
\textbf{Coarse-grained description of the 1stm coat protein.}
(\textit{a}) A view of the 1stm monomer created using
RasMol~\citep{Sayle:1995}.  The protein consists of more than
4,000 atoms in 157 amino acids.  (\textit{b}) The backbone of the
1stm protein. The color scale represents rigidity as determined by
the software FIRST with $\ecut = -0.7$ kcal/mol: Blue means rigid
and red means floppy while the other colors represent intermediate
rigidities~\cite{Jacobs:2001}. Adjacent amino acids with equal
binary rigidity are grouped into rigid or floppy domains, leading
to the typical pattern in capsid proteins: floppy ends and rigid
domains separated by short floppy domains at the center.
(\textit{c}) Schematic representation of the domain structure in
(\textit{b}). Here, rigid domains are drawn as rectangular
vertices and floppy domains are drawn as ellipses. The numbers
represent the number of amino acids in each domain. The thick
lines represent the covalent bonds in the backbone while the thin
lines are hydrogen bonds. Note that the domain structure is only
needed for the dissociation rates---the association rates can be
computed directly from the bond energies.

\subsubsection*{Figure~\ref{fig:1stmRigidityChange}}
\textbf{Change of rigidity as the assembly proceeds.} (\textit{a})
In $T=1$ viruses each cell in the icosahedral lattice
(Fig.~\ref{fig:1stmInterfaces}) accommodates one protein: here we
show a 1stm monomer in its reduced description, as in
Fig.{\ref{fig:1stmFlexibility}}(\emph{c}). As the assembly
proceeds, the rigidity of every oligomer is recalculated with
FIRST from the full atomic data. The existence of inter-molecular
hydrogen bonds in the 1stm dimer (\textit{b}) and trimer
(\textit{c}) modifies the rigidity of the constitutive monomeric
units. Monomers with the same color have identical rigidity
profile and domain partitioning.

\subsubsection*{Figure~\ref{fig:1stmAssemblyTree}}
\textbf{The combinatorial assembly tree of 1stm.} A tree
representing all the possible monomers, dimers and trimers in the
assembly of 1stm. The thin lines are possible assembly paths by
the addition of one monomer at a time. The thick lines represent
the pathways that are actually observed in the simulations
(compare with Fig.~\protect{\ref{fig:1stmReactionCartoon}}). The
number of possible pathways explodes combinatorially as the size
of the oligomer grows.

\subsubsection*{Figure~\ref{fig:1stmConcEA}}
\textbf{Simulations of the quasi-stationary state of 1stm lumped by
oligomer size.} (\textit{a}) Average time-weighted concentration of
the different oligomer sizes obtained in the Gillespie simulation
(crosses). The solid line is a guide to the eye. Note the high
concentration of dimers and hexamers. The circles (and dashed line)
show the prediction from the quasi-stationary Markov process
(Eq.~\protect{\ref{eq:QSD}}), which shows good agreement with the
simulation. (\textit{b}) Average time-weighted association energy per
monomer of the different oligomer sizes. The error bars are hardly
visible indicating that all different conformations of a given size
have almost identical association energies. Hexamers lie at a local
minimum, a clear signal that they are stable oligomers, relatively
more favorable than heptamers and octamers. Note the missing data
points for size 11, as no 11-mers are observed in the simulations.

\subsubsection*{Figure~\ref{fig:1stmPoissonGaussian}}
\textbf{Different concentration distributions in the simulations
of the 1stm assembly.} (\textit{a}) Most oligomers are present at
very low concentrations with Poisson-like distributions as
exemplified by the 1stm tetramers. This means that low
concentration is connected with short persistence. (\textit{b}) A
few oligomers have significant concentrations at all times in the
quasi-steady state, such as the 1stm dimers and hexamers shown.
Their distributions have Gaussian-like characteristics. Although
there are three distinct 1stm dimers, almost all dimers in the
simulation are of the most energetically favorable type (the one
circled in Fig.~\ref{fig:1stmReactionCartoon}).

\subsubsection*{Figure~\ref{fig:1stmReactionCartoon}}
\textbf{Stochastic sampling of the assembly pathways for 1stm with
$\geometryfactor = 10^{-4}$.} (\textit{a}) The heat map
illustrates the observed frequency of the reactions in the
assembly: white squares indicate no reactions involving the two
sizes, while darker shades indicate many reactions of that type.
Reactions above the diagonal represent associations ($M_i + M_j
\to M_k$, where $k=i+j$), while reactions below the diagonal
correspond to dissociations ($M_k \to M_i + M_j$, with $k=i+j$).
For example, the $(1,1)$ square in the top left corner represents
a `monomer plus monomer' association reaction, and the square to
the right $(1,2)$ is the association of monomer plus dimer.
Meanwhile, the $(2,1)$ square represents dimers splitting into two
monomers. (\textit{b}) Using the heat map in (\textit{a}) we
represent the most common oligomers of size six or less and the
transitions between them. Dashed arrows represent dissociation
reactions. Only reactions with a frequency above a given threshold
are represented. The thickness of the arrows is proportional to
the logarithm of the frequency of the reaction, which means that
the vast majority of the reactions involve forming or breaking up
oligomers of size six or smaller. The majority of reactions in the
system merge monomers to form dimers. Most hexamers are formed by
merging three dimers. But there is also a second pathway where a
trimer and a dimer form a pentamer that later adds a monomer to
complete the hexamer.(\textit{c}) A lumped, schematic
representation of the pathways in (\textit{b}) (inside the dotted
rectangle) showing also the higher steps of the cascade.

\subsubsection*{Figure~\ref{fig:4sbvConcEA}}
\textbf{Simulations of the quasi-stationary state of 4sbv lumped
by oligomer size.} (\textit{a}) Average time-weighted
concentration of the different oligomer sizes. There are peaks for
trimers, hexamers and other multiples of three indicating that the
trimer is an important building block in the assembly. Compare
with 1stm in Fig.~\ref{fig:1stmConcEA} (\textit{b}) Average
time-weighted association energy per monomer of the different
oligomer sizes. The error bars are hardly visible because all
different conformations of a given size have almost identical
association energies. Oligomers formed by multiples of three tend
to occupy local minima of the energy, a hint of their enhanced
stability. Note the missing data points for sizes 8, 14, 17, 19
and 20 since these oligomers are never observed in the
simulations.

\subsubsection*{Figure~\ref{fig:4sbvHeatMap}}
\textbf{Stochastic sampling of the assembly pathways for 4sbv with
$\geometryfactor = 2 \cdot 10^{-5}$.} (\textit{a}) The heat map
shows the frequency of the reactions taking place. Note how the
pattern differs from that of 1stm
(Fig.~\protect{\ref{fig:1stmReactionCartoon}}
and~\protect{~\ref{fig:1stmHeatMaps}}). There is a variety of
large oligomers present at any given time in the system and the
cascading behavior is less pronounced. The checkerboard pattern
indicates that reactions involving multiple-trimer oligomers are
the most common. (\textit{b}) A schematic view of the most common
reactions for 4sbv deduced from the heat map in (\textit{a}).
Again, note the differences with the assembly of 1stm
(Fig.~\protect{\ref{fig:1stmReactionCartoon}}).

\subsubsection*{Figure~\ref{fig:1stmGeometryFactor}}
\textbf{Impact of the form factor \geometryfactor\ on the speed of
assembly.} The figure shows the average proportion of dissociation
events for 1stm ($\times$) and 4sbv ($\circ$) calculated over 350
independent runs, with the standard deviation indicated by the
error bars. As $\kappa$ increases (making the association more
likely), the proportion of dissociation events decays from $0.5$
towards zero. Therefore, $\kappa$ is directly related to the
overall speed at which the assembly proceeds. If $\kappa$ is
small, the cascading process is not initiated and the assembly
stalls. Clearly, detailed models for the calculation of the form
factor $\kappa$ would be of importance. Note the different
$\kappa$ values at which both viruses would start to assemble,
which is reminiscent of their proclivity to self-assembly \emph{in
vitro.}

\subsubsection*{Figure~\ref{fig:1stmHeatMaps}}
\textbf{Impact of the form factor \geometryfactor\ on the pathways
of assembly.} Heat maps depicting the pattern of reactions for
1stm with (\textit{a}) $\geometryfactor = 10^{-3}$ and
(\textit{b}) $\geometryfactor = 10^{-5}$. As compared to
Fig.~\protect{\ref{fig:1stmReactionCartoon}}, note that the
overall cascading pattern of reactions and pathways remains
broadly unchanged. In (\textit{a}), larger \geometryfactor\ means
that there are fewer dissociation events, i.e., fewer dark squares
below the diagonal, see Fig.~\ref{fig:1stmGeometryFactor}. In
(\textit{b}), smaller $\kappa$ means more dissociation events, but
the pattern of association for the larger oligomers is similar for
all three 1stm heat maps.

\subsubsection*{Figure~\ref{fig:1stmCapsidCompletion}}
\textbf{Kinetics of the completion of the full capsid for 1stm.}
(\textit{a}) Average concentration of all oligomer sizes as a
function of time. At $t=0$ all proteins are monomers and they
rapidly react to form larger oligomers. The most conspicuous
feature of the distribution is that there are very few oligomers
of intermediate sizes. (\textit{b}) Concentration of the most
common oligomers: 1-- (solid line), 2-- (dashed line), 6--
($\times$) and 60--mers ($\circ$) as a function of time. All other
oligomers have negligible concentrations. (\textit{c}) Time for
the completion of SPMV capsids with $\geometryfactor = 10^{-3}$.
For this value of \geometryfactor, the number of dissociation
events is almost negligible, as shown in
Fig.~\ref{fig:1stmGeometryFactor}.

\begin{figure}
  \begin{center}
    \includegraphics[width = 3.25in]{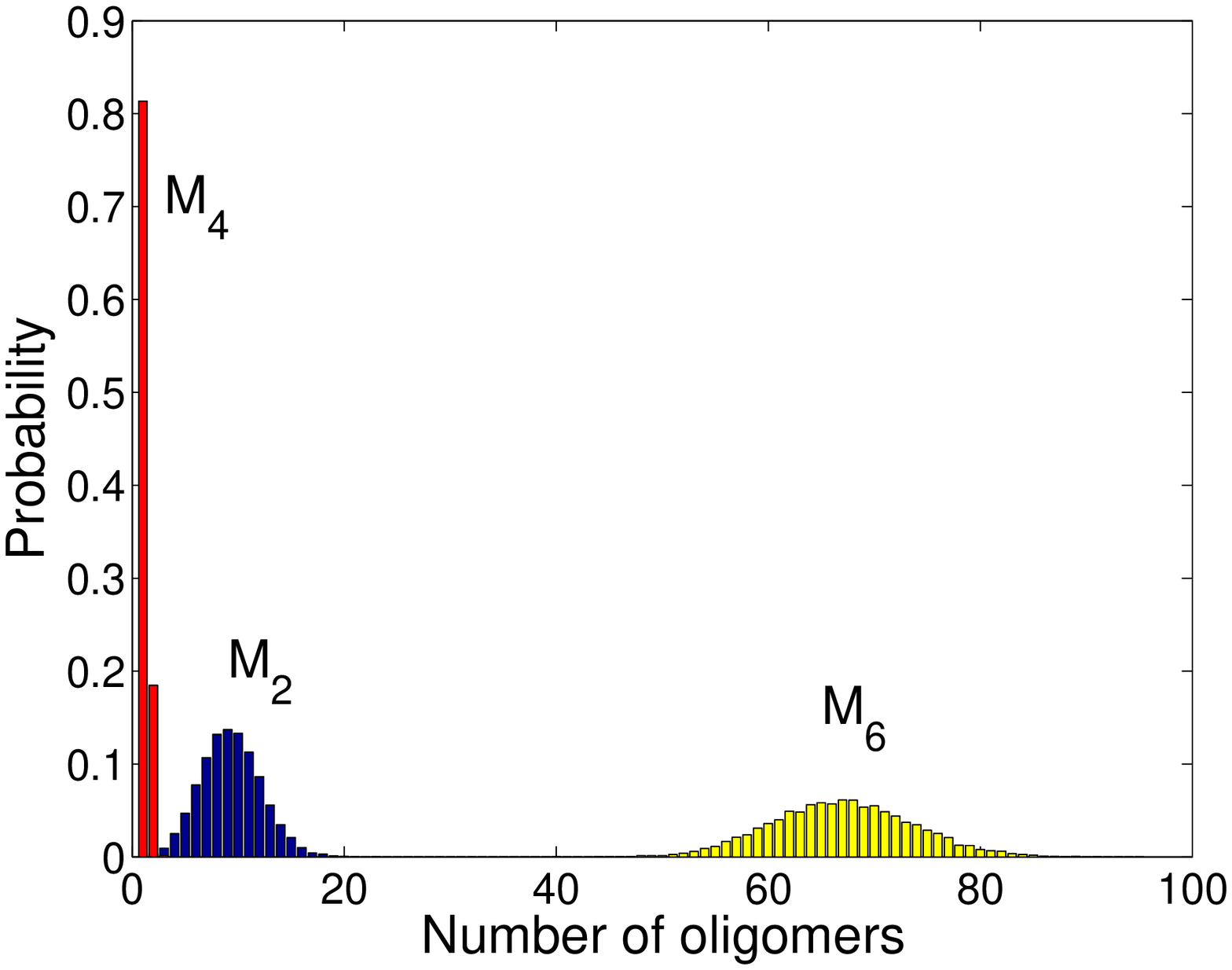}
    \caption{}
    \label{fig:1stmToySystem}
  \end{center}
\end{figure}

\subsubsection*{Figure~\ref{fig:1stmToySystem}}
\begin{figure}
  \begin{center}
    \includegraphics[width = 3.25in]{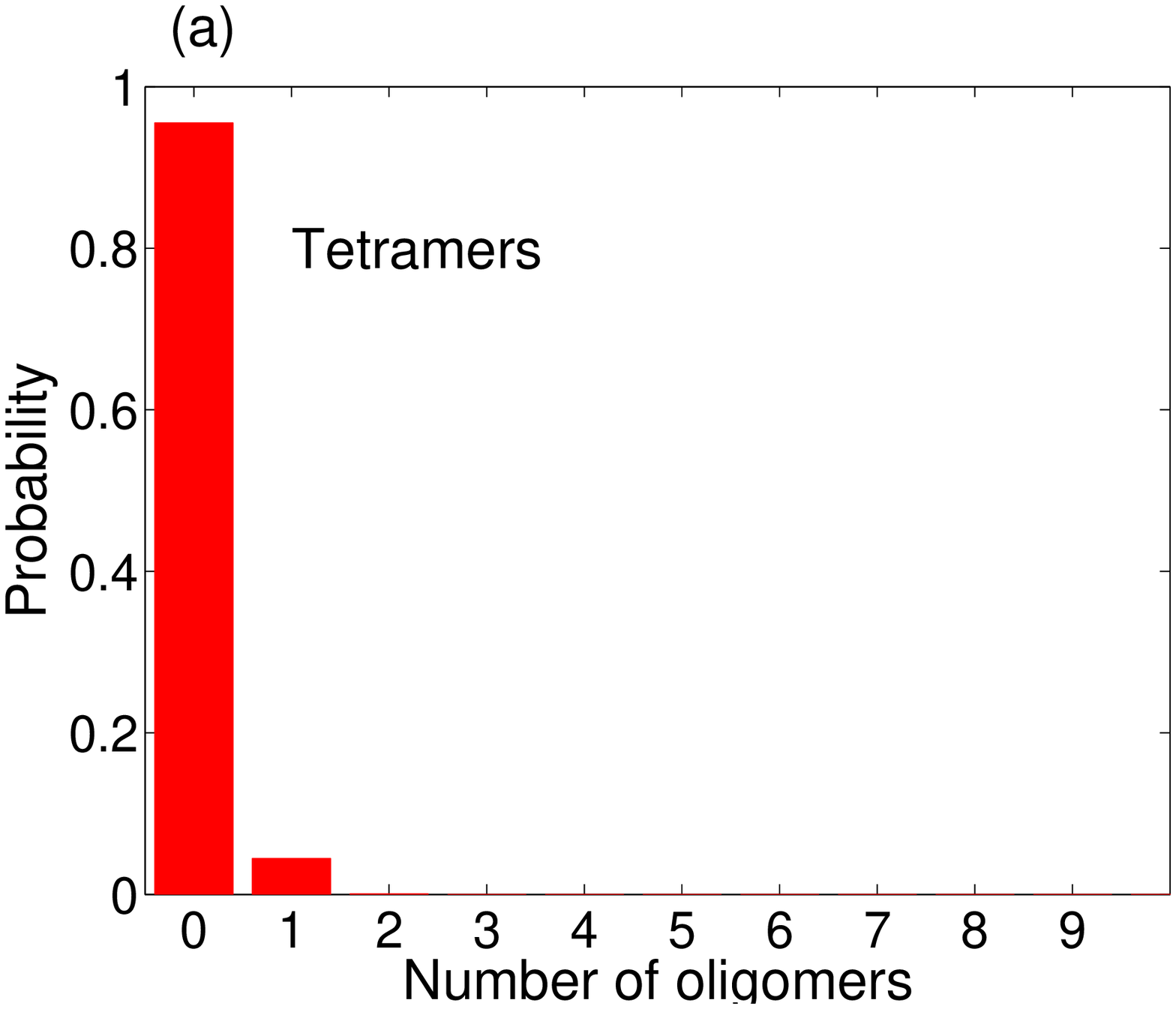}\\
    \includegraphics[width = 3.25in]{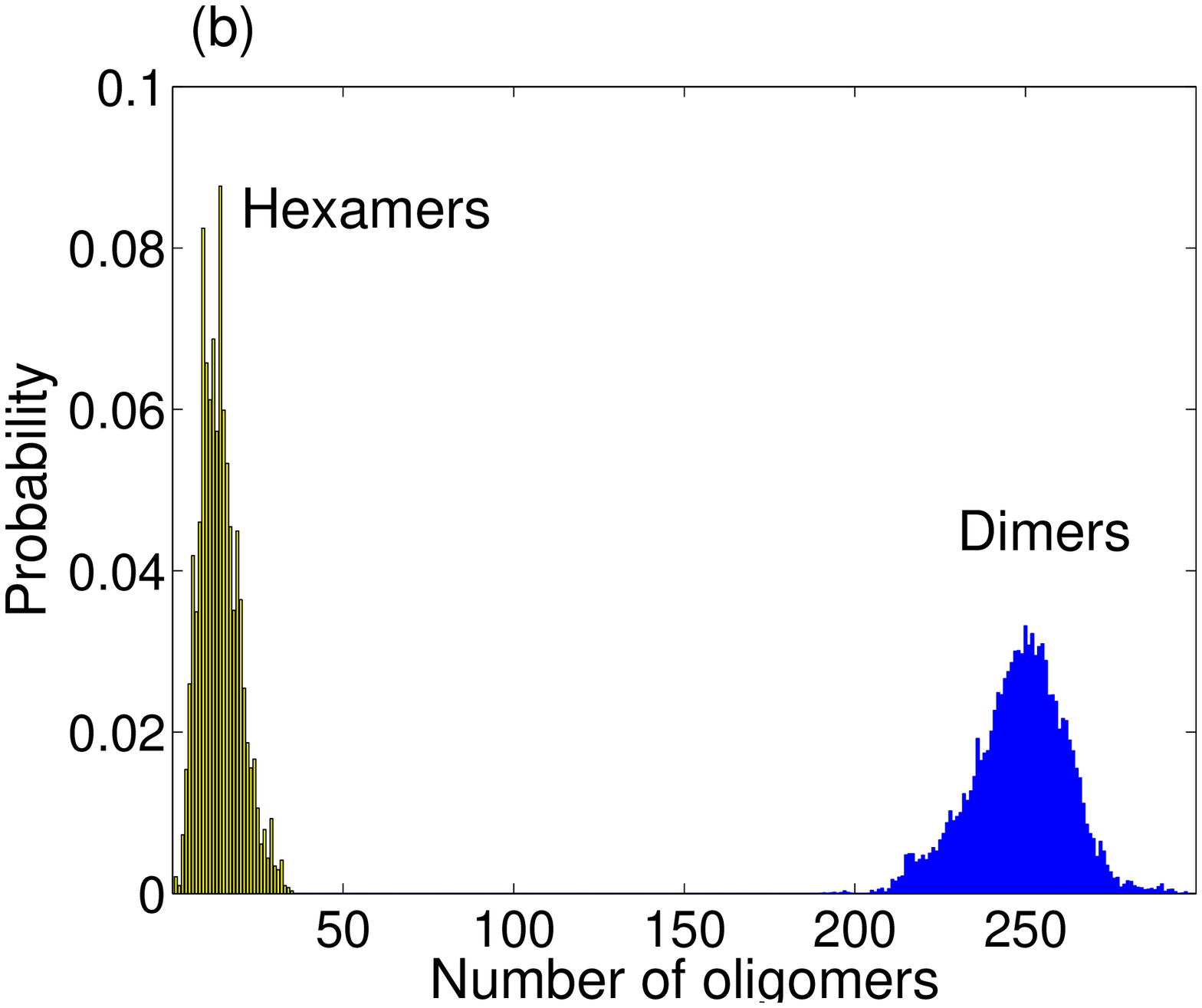}
    \caption{}
    \label{fig:1stmPoissonGaussian}
  \end{center}
\end{figure}

\textbf{Different oligomer distributions in a simple cascading
system.} The simple cascading reaction
Eq.~\protect{\ref{eq:simple_reaction}} involves only three
different types of oligomers: `dimers' ($M_2$), `tetramers'
($M_4$) and `hexamers' ($M_6$). In the quasi-steady state,
`dimers' and `hexamers' have Gaussian-like distributions while the
distribution of `tetramers' is Poisson-like (compare with
Fig.~\protect{\ref{fig:1stmPoissonGaussian}}).

\clearpage

\clearpage

\clearpage

\clearpage

\clearpage

\clearpage

\clearpage

\clearpage

\clearpage

\clearpage

\clearpage

\end{document}